\documentclass{article}

\usepackage[english]{babel}
\usepackage[utf8]{inputenc}
\usepackage{natbib}
\usepackage{csquotes}
\usepackage{soul}
\usepackage{comment}
\usepackage{longtable}
\usepackage{booktabs}
\usepackage{array}
\usepackage{lipsum}
\usepackage[margin=1in]{geometry}
\usepackage[bottom]{footmisc}

\usepackage{tablefootnote}
\usepackage{definitions} %import definitions.sty
\usepackage{pdflscape}
\usepackage{multirow}
\usepackage{footnote}

\makesavenoteenv{tabular}

\usepackage{graphicx}
\usepackage{caption}
\usepackage{subcaption}

\usepackage{float}
\usepackage{amsmath}
\usepackage{bm}
\usepackage{amssymb}

\usepackage{amsthm}
\usepackage{mathtools}
\usepackage{graphicx}
\usepackage{mathtools}
\usepackage{makecell}

\usepackage{scalerel}
\usepackage{tikz}
\usetikzlibrary{svg.path}

\definecolor{orcidlogocol}{HTML}{A6CE39}
\tikzset{
  orcidlogo/.pic={
    \fill[orcidlogocol] svg{M256,128c0,70.7-57.3,128-128,128C57.3,256,0,198.7,0,128C0,57.3,57.3,0,128,0C198.7,0,256,57.3,256,128z};
    \fill[white] svg{M86.3,186.2H70.9V79.1h15.4v48.4V186.2z}
                 svg{M108.9,79.1h41.6c39.6,0,57,28.3,57,53.6c0,27.5-21.5,53.6-56.8,53.6h-41.8V79.1z M124.3,172.4h24.5c34.9,0,42.9-26.5,42.9-39.7c0-21.5-13.7-39.7-43.7-39.7h-23.7V172.4z}
                 svg{M88.7,56.8c0,5.5-4.5,10.1-10.1,10.1c-5.6,0-10.1-4.6-10.1-10.1c0-5.6,4.5-10.1,10.1-10.1C84.2,46.7,88.7,51.3,88.7,56.8z};
  }
}

\newcommand\orcidicon[1]{\href{https://orcid.org/#1}{\mbox{\scalerel*{
\begin{tikzpicture}[yscale=-1,transform shape]
\pic{orcidlogo};
\end{tikzpicture}
}{|}}}}

\usepackage{hyperref}

\newcolumntype{x}[1]{>{\centering\arraybackslash}p{#1}}

\usepackage{xargs}
\usepackage[colorinlistoftodos,prependcaption,textsize=tiny]{todonotes}
\newcommandx{\unsure}[2][1=]{\todo[linecolor=red,backgroundcolor=red!25,bordercolor=red,#1]{#2}}
\newcommandx{\change}[2][1=]{\todo[linecolor=blue,backgroundcolor=blue!25,bordercolor=blue,#1]{#2}}
\newcommandx{\info}[2][1=]{\todo[linecolor=OliveGreen,backgroundcolor=OliveGreen!25,bordercolor=OliveGreen,#1]{#2}}
\newcommandx{\improvement}[2][1=]{\todo[linecolor=Plum,backgroundcolor=Plum!25,bordercolor=Plum,#1]{#2}}

\DeclareMathOperator{\Ber}{Ber}

\DeclareMathOperator{\Beta}{Beta}
\usepackage{bbm}

\title{Joint Gaussian Graphical Model Estimation: A Survey}
\author{Katherine Tsai$^1$\,\orcidicon{0000-0002-7243-5260},\;Oluwasanmi Koyejo$^2$\,\orcidicon{0000-0002-4023-419X
},\;Mladen Kolar$^3$\;\orcidicon{0000-0001-7353-3404} }

\date{%
    $^1$ Department of Electrical and Computer Engineering, University of Illinois at Urbana-Champaign\\%
    $^2$ Department of Computer Science, University of Illinois at Urbana-Champaign\\%
    $^3$ The University of Chicago Booth School of Business\\[2ex]%
}

\begin{document}

\maketitle
{\bf Conflict of interest} The authors declare that there is no conflict of interest. 

{\bf Article type} Advanced review 

{\bf Correspondence} Mladen Kolar, The University of Chicago Booth School of Business, Chicago, IL, USA. Email: mladen.kolar@chicagobooth.edu

\section*{Abstract}

Graphs representing complex systems often share a partial underlying structure across domains while retaining individual features. Thus, identifying common structures can shed light on the underlying signal, for instance, when applied to scientific discovery or clinical diagnoses. Furthermore, growing evidence shows that the shared structure across domains boosts the estimation power of graphs, particularly for high-dimensional data.  However, building a joint estimator to extract the common structure may be more complicated than it seems, most often due to data heterogeneity across sources. This manuscript surveys recent work on statistical inference of joint Gaussian graphical models, identifying model structures that fit various data generation processes.\\ 
{\bf Keywords}: Gaussian graphical model; joint network; graphical lasso;  high-dimensional estimation; sparsity 
\begin{figure}[h]
    \centering
    \includegraphics[width=.5\textwidth]{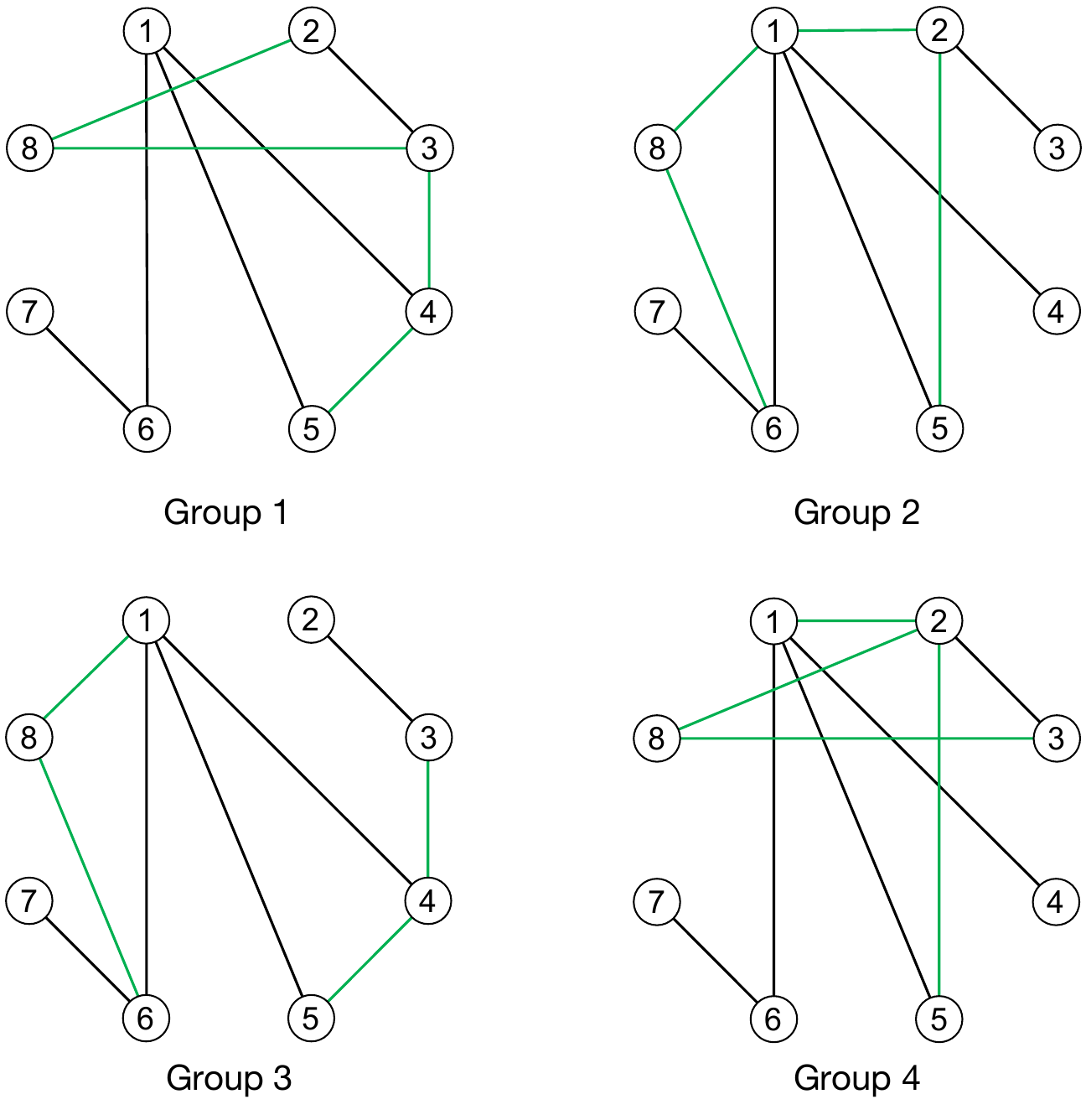}
    \caption{{\bf Graphical Abstract}. Joint graphical model estimation studies a group of graphs that have partially shared edge structures, presented in black, and individually owned edge structures, presented in green. Jointly estimating the shared structures enhances the estimation power while preserving individual structures as well.}
    \label{fig:my_label}
\end{figure}

\section{Introduction}\label{s:intro}

Graphical models are powerful tools for expressing statistical relationships between variables. Examples of practical uses are ubiquitous and include models that characterize the causal relationships between the neurological activity of brain regions,  genetic expression across genes, and a variety of other physiological measurements. A variety of applications have illustrated the value of graphical models for analyzing scientific phenomena~\citep{felsenstein1981evolutionary,schafer2005shrinkage, friedman2000using, chan2017gene, dondelinger2013non}. Specifically, graphical models have proven useful for elucidating the mechanisms of brain function~\citep{foti2019statistical,manning2018probabilistic, schwab2018directed, greenewald2017time, colclough2018multi, qiu2016joint, SKRIPNIKOV2019164}. This manuscript outlines joint graphical models, an extension to standard graphical models that are useful for jointly analyzing data from multiple sources, e.g., neurological data measured at multiple timescales, or joint neurological, genetic and phenotypic data. Specifically, this manuscript lays out the representation of joint graphical models and some of their properties, then outlines the best practices for estimating joint graphical models. This manuscript provides examples of data generation processes where the joint approach can significantly improve estimates compared to separate estimation.\\\par%

A graph $G = (V,E)$ consists of a set of $p$ nodes, also known as vertices $V = \{1, \ldots, p\}$ and a set of edges $E \subseteq V \times V$. In a probabilistic graphical model~\citep{lauritzen1996graphical}, the set of nodes $V$ is associated with coordinates of a random vector ${\bf x} = (x_1, \ldots, x_p)^\top$ and the edge set $E$ captures dependency relationships between the components of the vector. In particular, in an undirected probabilistic graphical model, the absence of an edge between nodes $a$ and $b$ indicates that $x_a$ and $x_b$ are conditionally independent given all other variables ${\bf x}_{-\{a,b\}} = \{x_c \mid c \in V\backslash\{a,b\} \}$.  {In the case when $a$ is a subset of nodes $\{1,\ldots,p\}$ rather than a single node, we will denote ${\bf x}_a\in\mathbb{R}^{|a|}$ as the vector whose entries correspond to $x_i$ for $i\in a\subseteq\{1,\ldots,p\}$.} In a neuroscience application, the random vector ${\bf x}$ could represent, for example, measurements of brain activity in different regions -- so the set of edges corresponds to functional brain connectivity. Given $n$ measurements of the vector ${\bf x}$, inferring the graph structure corresponds to identifying pairs of coordinates that are conditionally independent given all other variables~\citep{Drton2016Structure}. Inferring the graph structure based on conditional associations is more challenging than inferring the correlation structure between the measurements. However, the conditional independence graphs are generally considered more scientifically meaningful~\citep{dobra2004sparse}. \\\par

\section{Background: Gaussian Graphical Models}

The most widely used examples of probabilistic graphical models are Gaussian graphical models, where ${\bf x} \sim \mathcal{N}({\bm\mu}, {\bm\Omega}^{-1})$ is assumed to be distributed as a multivariate Gaussian vector with the mean vector ${\bm\mu}$ and the precision matrix ${\bm\Omega}$ whose entries correspond to the partial correlation between the associated variables. In this setting, any two coordinates $x_a$ and $x_b$ are conditionally independent given ${\bf x}_{-\{a,b\}}$ if and only if the $(a,b)$ entry of the precision matrix ${\bm\Omega}$ is zero~\citep{lauritzen1996graphical}, and the graph structure can be inferred based on nonzero entries of ${{\bm\Sigma}}^{-1}:={\bm\Omega}$, also known as the inverse covariance matrix. Throughout the manuscript, we use the terms inverse covariance matrix and precision matrix interchangeably. In practice, the covariance matrix is not known and the graph structure needs to be estimated using samples drawn from an underlying distribution. For example, in a low-dimensional setting, we can first obtain an estimator of the precision matrix by maximizing the log-likelihood 
\begin{align}\label{eq:MLLN}
    \hat{\bm\Omega}=\argmax\quad n\left[\frac{1}{2}\log\{\det({\bm\Omega})\}-\frac{1}{2}\tr(\hat{\bm\Sigma}{\bm\Omega})\right],
\end{align}
where $\det(\cdot)$ is the determinant, $\tr(\cdot)$ is the trace, ${\bf x}=n^{-1}\sum_{i=1}^n{\bf x}_i$ is the empirical mean and  $\hat{{\bm\Sigma}} = n^{-1}\sum_{i=1}^n({\bf x}_i - \bar {\bf x})({\bf x}_i - \bar {\bf x})^\top$ is the empirical covariance matrix. Next, the graph structure is estimated by thresholding small (in absolute value) elements of $\hat{{\bm\Omega}}$ or testing whether they are zero~\citep{drton2004model, Drton2016Structure}, that is, the graph structure corresponds to the nonzero entries of the resulting thresholded precision matrix.\\\par

\begin{figure}[t]
    \centering
    \includegraphics[width=0.7\textwidth]{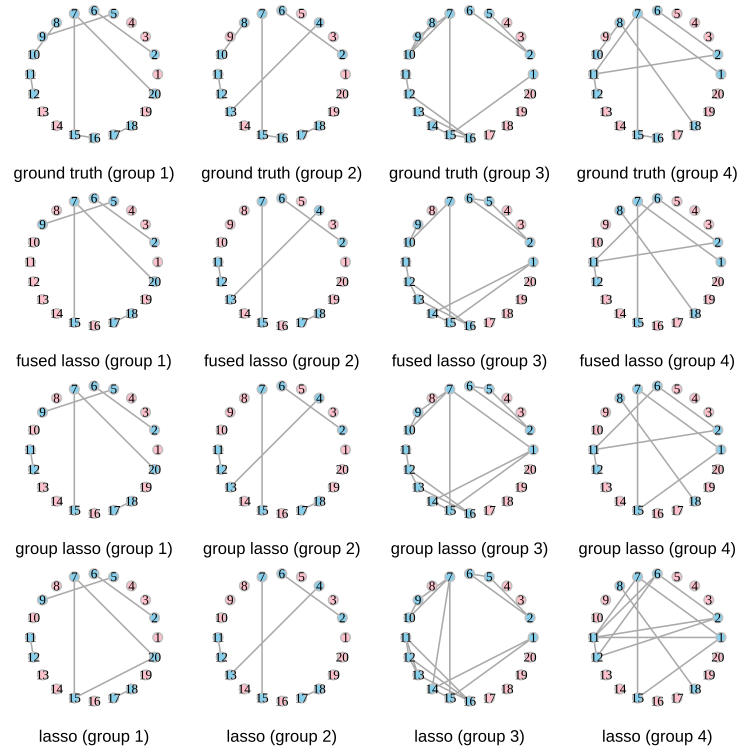}
\caption{Comparison of joint and separate graph estimation procedures with $p=20$ variables and sample size of $n=5000$. Pink nodes do not have edges connecting to them, while blue nodes have at least on edge connecting to them. {\bf (Top row):} Ground truth graphs. {\bf (Second row):} Jointly estimated graphs using fused lasso. {\bf (Third row):} Jointly     estimated graphs using the group graphical lasso. {\bf (Fourth row):} Graphs estimated separately using the graphical lasso. Details of all methods are provided in the text. From the figure, it is clear that joint estimation significantly outperforms separate estimation.}
    \label{fig:graph_visualize}
\end{figure}

In a high-dimensional setting, where the number of parameters to estimate, $p$, is much larger than the number of data points observed, $n$, maximizing the log-likelihood~\eqref{eq:MLLN} results in poor quality estimates. In the particular case of Gaussian data, the resulting estimate, that is, the inverse of the covariance matrix, does not exist when $n < p$. Unfortunately, the high-dimensional setting is prevalent in various applications. For example, functional imaging of brain measurements using (standard) $2 mm^3$ voxels will result in approximately $p=O(10^6)$ voxels with $n = O(10^2)$ measurements~\citep{poldrack2011handbook, hsieh2013big}. {There are two common problems that neuroscientists are interested in studying: (i) a static graph representing conditional independence between time series~\citep{foti2019statistical} and (ii) time-varying graphs within individuals~\citep{calhoun2014chronnectome, lurie2020questions}. In the first problem, we estimate a single graph by treating each time point as an i.i.d.~sample~\citep{varoquaux2010detection}, effectively ignoring the temporal dependence. In the second problem, we estimate graphs for different time points or graphs within a time window. We will cover associated methodologies for both problems in Section~\ref{section:coarse-grained_structure} -- \ref{section:fine-grained_structure} and Section~\ref{section:time_series_data}, respectively. We note that in addition to the small sample sizes, applications to fMRI are affected by temporal correlations in the observed data, which can reduce the effective sample size~\citep{qiu2016joint}.} While $p$ is large as compared to $n$, most entries in ${\bm\Omega}$, denoted as $\omega_{i,j}$, $i,j=1,\ldots,p$, are zero, that is, the inverse covariance matrix is sparse. Hence, a typical strategy to estimate ${\bm\Omega}$ in a high-dimensional setting is to add a regularization function, such as the $\ell_1$-norm of the parameters, to the log-likelihood function~\eqref{eq:MLLN}, which encourages the graph to be sparse or have other desirable structural biases~\citep{yuan2007model,buhlmann2011statistics}. {Specifically, we estimate ${\bm\Omega}$ using the following optimization program
\begin{align}\label{penalizedMLE}
    \hat{\bm\Omega}=\argmax\quad n\left[\frac{1}{2}\log\{\det({\bm\Omega})\}-\frac{1}{2}\tr(\hat{\bm\Sigma}{\bm\Omega})\right]
    -
    \lambda_n\sum_{i\neq j}|\omega_{i,j}|.
\end{align}}
Note that regularization is not added to the diagonal terms, $\omega_{i,i}$, $i=1,\ldots,p$, because ${\bf\Omega}$ is positive definite and adding penalty on the diagonal entries would introduce additional bias. In this manuscript, we focus on the simultaneous estimation of multiple graphs that are structurally similar. We will illustrate in the following sections that exploiting the common structures will improve the estimation results on every individual graph.

\subsection{Joint Gaussian Graphical Models}\label{section:general_framework}

We continue to use the example of brain measurements to demonstrate the idea of joint graphical model estimation. Consider the case of $n$ fMRI scans collected from each of the $K$ subjects. Suppose that we seek to estimate the functional connectivity (in this case, a graphical model) between the $p$ voxels of each subject $k$, where $p$ is much greater than $n$ ($p\gg n$). Each functional connectivity network shares similarities with other networks, but are not identical. To better estimate the network of the subject $k$ from $n$ MRI scans, we may borrow information from other networks given that they are expected to share similar patterns. One simple approach is to construct a regularization function that encourages similarities between graphs, an approach known as joint estimation. Figure~\ref{fig:graph_visualize} shows a promising result when one pools the data across subjects, the resulting estimates better recover the ground truth graphs compared to separate estimations. It has also been shown that joint estimation can increase sensitivity and detect edges that are missing in separate estimation~\citep{chiquet2011inferring, peterson2015bayesian}. Thus, ignoring the information of other groups may lead to suboptimal solutions~\citep{danaher2014joint, lee2015joint}. Moreover, joint estimation of graphical models has been applied successfully in a number of problems, including metabolite experiments~\citep{tan2017bayesian}, cancer networks~\citep{mohan2012structured,peterson2015bayesian,lee2015joint,saegusa2016joint,hao2017simultaneous}, biomedical data~\citep{yajima2014detecting,kling2015efficient, pierson2015sharing}, gene expression~\citep{ chun2015gene, lin2017joint}, text processing~\citep{guo2011joint}, climate data~\citep{ma2016joint}, and fMRI \citep{ qiu2016joint, colclough2018multi, SKRIPNIKOV2019164, lukemire2020bayesian}. In all of these problems, data are heterogeneous, but the graphs share similarities.\\\par

To rigorously describe the example discussed above, we consider the problem of estimating graph structures $G^{(k)}=(V, E^{(k)})$, $k=1,\hdots,K$, from $K$ related groups of data. The data for each group are $p$-variate and share the same set of nodes $V$, but the underlying connection patterns $E^{(k)}$ may be different due to the heterogeneity between groups.  The data for the $k$-th group can be represented as an $n_k\times p$ matrix ${\bf X}^{(k)}=({\bf x}_1^{(k)}, \hdots, {\bf x}_{n_k}^{(k)})^\top$, where the rows ${\bf x}_{i}^{(k)}=(x_{i,1}^{(k)},\hdots,x_{i,p}^{(k)})$, $i=1,\hdots,n_k$, are $p$-dimensional vectors of observations.  Assuming that the data in each group are distributed according to a $p$-variate Gaussian distribution,  ${\bf x}_{i}^{(k)}\sim\mathcal{N}({\bm\mu}^{(k)}, \{{\bm\Omega}^{(k)}\}^{-1})$, $i=1,\hdots,n_k$, where ${\bm\mu}^{(k)}\in\mathbb{R}^p$ is the mean, which we assume without loss of generality to be $\bm0$, and ${\bm \Omega}^{(k)}\in\mathbb{R}^{p\times p}$ is the precision matrix.
{Given  observations ${\bf X}=\{{\bf X}^{(1)},\hdots,{\bf X}^{(K)}\}$, we can estimate ${\bm\Omega}=\{{\bm\Omega}^{(1)},\hdots,{\bm\Omega}^{(K)}\}$ by maximizing the penalized joint log-likelihood for $K$ groups:
\begin{align}\label{jointMLE}
    \hat{{\bm\Omega}}=\argmax\;{\ell}({\bm \Omega})
    -
    {\bf P}({\bm\Omega}),
    \qquad {\ell}({\bm \Omega}) :=\sum_{k=1}^Kn_k\left[\log\{\det({\bm\Omega}^{(k)})\}-\tr(\hat{{\bm\Sigma}}^{(k)}{\bm\Omega}^{(k)})\right],
\end{align}
where $\hat{{\bm\Sigma}}^{(k)}={n_k}^{-1}({\bm X}^{(k)})^\top{\bm X}^{(k)}$, $k=1,\hdots,K$, are the sample covariance matrices.} Directly solving~\eqref{jointMLE} without the penalty ${\bf P}({\bm\Omega})$ gives the maximum log-likelihood estimate of ${\bm\Omega}$. However, the solution is equivalent to solving the maximum  log-likelihood estimate of each group individually and fails to utilize the shared ``{information}'' across different groups. We hence explore different approaches that use the penalty function ${\bf P}({\bm\Omega})$ to incorporate the group structure and focus on the structural assumptions behind the penalties. The comparison of different methods introduced in the text can be found in Table~\ref{tab:summary}. {Specifically, we consider coarse-grained vs.~fine-grained structural assumptions. For coarse-grained structures, all pairs of edge strengths are penalized/regularized in the same way, i.e., invariant to the group identity. In contrast, fine-grained structure uses regularization/priors between edge strengths that vary across groups, e.g., using prespecified weights for pairs of groups. }
\\\par
%{add discussion about joint thresholding of multiple graphical models}\\\par
The rest of the manuscript is organized as follows. In Section~\ref{section:coarse-grained_structure}, we introduce methods that employ coarse-grained structural constraints. Methods that employ fine-grained structural constraints are discussed in Section~\ref{section:fine-grained_structure}. Section~\ref{section:differential_graph} and \ref{section:time_series_data} illustrate two practical examples. Section~\ref{section:differential_graph} covers differential graphs, which are special cases of the joint estimation paradigm with two groups.  Joint estimation of time-series data is discussed in Section~\ref{section:time_series_data}. Finally, we close the review with open problems in Section~\ref{section:conclusion}.

%Finally, in Section~\ref{section:experiments}, we present an empirical evaluation of a variety of
%regularization methods to evaluate their performance under two graph
%generation processes: shared inverse covariance structure and shared
%inverse covariance values. The simulation code
%is available at
%\url{https://github.com/koyejo-lab/JointGraphicalLasso}.
\newpage
\begin{landscape}
\centering
\renewcommand*{\arraystretch}{1.2}
\begin{longtable}{|m{1.5cm}|m{2.5cm}|m{4cm}|m{5.5cm}|m{8.5cm}|}
\caption{Variants of the joint Gaussian graphical model \label{tab:summary}}\\
%\fontsize{8pt}{8pt}\selectfont
\hline
Category& Method& Model Name & Model Structure  ${\bm\Omega}$&\multicolumn{1}{l|}{Penalty Function/Negative Log Prior/Constraint}\\
\hline
\multirow{28}{*}{\shortstack[l]{Coarse\\-grained}}& \multirow{28}{*}{\shortstack[l]{Penalized\\ MLE}}
&JGL~\citep{guo2011joint} -- Section~\ref{ssec:HSN}& \multirow{2}{*}{$\left\{\begin{array}[c]{@{}l@{}}\omega_{i,j}^{(k)}=\theta_{i,j}\gamma_{i,j}^{(k)},\;i\neq j;\\ \omega_{i,i}^{(k)} = \gamma_{i,i}^{(k)},\;i=j.\end{array}\right.$}&
{$\lambda_1\sum_{i\neq j}\theta_{i,j}+\lambda_2\sum_{k=1}^{K}\sum_{i\neq j}|\gamma_{i,j}^{(k)}|$}\\
&&&&\\\cline{3-5}

&& JWLGL~\citep{shan2020joint} -- Section~\ref{ssec:HSN}&
\multirow{5}{*}{\shortstack[l]{$\omega_{i,j}^{(k)}=\theta_{m,m'}^{(k)}\gamma_{i,j}^{m,m',(k)}.$\\
$
\theta_{m,m'}^{(k)}=\left\{\begin{array}{cc}
    \alpha_{m,m'}\beta_{m,m'}^{(k)}, &  m\neq m';\\
    1, & \text{otherwise}.
\end{array}\right.
$\\
$\gamma_{i,j}^{m,m,(k)} = \left\{\begin{array}{ll}
        \iota_{i,j}^{m,m}\rho_{i,j}^{m,m,(k)}  ,& i\neq j; \\
        1 ,& i=j.
    \end{array}\right.$}}
&{
\multirow{4}{*}{\shortstack[l]{
$\lambda_1\sum_{m\neq m'}\alpha_{m,m'}
$\\
$+ \lambda_2\sum_{m\neq m'}\sum_{k=1}^K\sum_{i\neq j}|\beta_{m,m'}^{(k)}\gamma_{i,j}^{m,m',(k)}|$\\
$+ \lambda_3 \sum_{m=1}^M\sum_{i\neq j}\iota_{i,j}^{m,m}$\\
$+\lambda_4\sum_{m=1}^M\sum_{i\neq j}\sum_{k=1}^K|\rho_{i,j}^{m,m,(k)}|
$
}}
}\\
&&&&\\
&&&&\\
&&&&\\
&&&&\\
&&&&\\\cline{3-5}
&& 
FGL~\citep{danaher2014joint} -- Section~\ref{ssec:coarse_reg}
&$\omega_{i,j}^{(k)}$&{$\lambda_1\sum_{k=1}^K\sum_{i\neq j}|\omega_{i,j}^{(k)}|+\lambda_2\sum_{k<k'}\sum_{i,j}|\omega_{i,j}^{(k)}-\omega_{i,j}^{(k')}|$}\\\cline{3-5}
&& GGL~\citep{danaher2014joint} -- Section~\ref{ssec:coarse_reg}
&$\omega_{i,j}^{(k)}$&{$\lambda_1\sum_{k=1}^K\sum_{i\neq j}|\omega_{i,j}^{(k)}|+\lambda_2\sum_{i\neq j}\cbr{\sum_{k=1}^K(\omega_{i,j}^{(k)})^2}^{\frac{1}{2}}$}\\\cline{3-5}
&&JAGL~\citep{shan2018joint} -- Section~\ref{ssec:coarse_reg}&$\omega_{i,j}^{(k)}$ & $\sum_{k=1}^K\frac{1}{n_k}\sum_{i\neq j}\frac{1}{|(1-\pi)\hat{t}_{i,j}+\pi\hat{s}_{i,j}^{(k)}|^r}|\omega_{i,j}^{(k)}|$\footnote{$n_k$ denotes the number of samples of group $k$, $\hat{t}_{i,j}$ is the precision matrix estimated by pooling all samples across groups, $\hat{s}_{i,j}$ is the precision matrix estimated by an individual group, and $r>0$}\\\cline{3-5}
&&TFRE~\citep{bilgrau2020targeted} -- Section~\ref{ssec:coarse_reg}&$\omega_{i,j}^{(k)}$
&\multirow{2}{*}{\shortstack[l]{$\sum_{k=1}^K\frac{\lambda_k}{2}\|{\bm\Omega}^{(k)}-{\bf T}^{(k)}\|_F^2$\\$+\sum_{k_1,k_2}^K\frac{\lambda_{k_1,k_2}}{4}\|({\bm\Omega}^{(k_1)}-{\bf T}^{(k_1)})-({\bm\Omega}^{(k_2)}-{\bf T}^{(k_2)})\|_F^2$}}\\
&&&&\\\cline{3-5}
&& SCAN~\citep{hao2017simultaneous} -- Section~\ref{ssec:coarse_reg}
&$\omega_{i,j}^{(k)}$&\multirow{2}{*}{\shortstack[l]{$\lambda_1\sum_{k=1}^K\sum_{i\neq j}|\omega_{i,j}^{(k)}|+\lambda_2\sum_{i\neq j}\left\{\sum_{k=1}^K(\omega_{i,j}^{(k)})^2\right\}^{\frac{1}{2}}$\\$+\lambda_3\sum_{k=1}\sum_{i=1}^p|\mu_i^{(k)}|$}}\\
&&&&\\
&&&&\\\cline{3-5}
&&RCON~\citep{mohan2012structured, mohan2014node} -- Section~\ref{ssec:diffnetwork_reg}&$\omega_{i,j}^{(k)}$&\multirow{2}{*}{\shortstack[l]{$G_q({\bm\Omega}^{(1)}-{\bm\Omega}^{(2)})=\min_{V:{\bm\Omega}^{(1)}-{\bm\Omega}^{(2)}=V+V^\top}f(V)$\\$f(V)=\sum_{j=1}^p\|V_j\|_q$}}\\
&&&&\\\cline{3-5}
&&
GFGL~\citep{gibberd2017regularized} -- Section~\ref{ssec:timeseries_reg}
&$\omega_{i,j}^{(k)}$&$\lambda_1\sum_{t=1}^T\sum_{i\neq j}|\omega_{i,j}^{(t)}|+\lambda_2\sum_{t=2}^T\|{\bm\Omega}^{(t)}_{-ii}-{\bf\Omega}_{-ii}^{(t-1)}\|_F$
\\\cline{2-5}
\pagebreak
\cline{2-5}
&\multirow{4}{*}{\shortstack[l]{CLIME\\
\citep{cai2011constrained}}}&JEMP~\citep{lee2015joint} -- Section~\ref{ssec:HSN}&$\omega_{i,j}^{(k)}=\theta_{i,j}+\gamma_{i,j}^{(k)}$&
\multirow{2}{*}{\shortstack[l]{
       $|\frac{1}{K}\sum_{k=1}^K\{\hat{{\bm\Sigma}}^{(k)}({\bm \Theta}+{\bm \Gamma}^{(k)})-{\bm I}\}|_{\infty}\leq\lambda_1$\\
    $|\hat{{\bm\Sigma}}^{(k)}({\bm \Theta}+{\bm \Gamma}^{(k)})-{\bm I}|_{\infty}\leq\lambda_2,\;\sum_{k=1}^K {\bm \Gamma}^{(k)}=0$}}\\
&&&&\\\cline{3-5}
&&KSE~\citep{qiu2016joint} -- Section~\ref{ssec:ksgm_timeseries}&$\omega_{i,j}^{(k)}$&$|\hat{{\bm S}}(u_0){\bm\Omega}(u_0)-{\bm I}|_\infty\leq\lambda_1\nonumber$\\\cline{2-5}
&\multirow{8}{*}{\shortstack[l]{Bayesian\\approach}}&DSS-JGL~\citep{li2019bayesian} -- Section~\ref{ssec:coarse_bayesian}
&$\omega_{i,j}^{(k)}$&\multirow{2}{*}{
\shortstack[l]{$\lambda_1\sum_{k=1}^K\sum_{i=1}^p
    |\omega_{i,i}^{(k)}|
    +
    \lambda_2\sum_{k=1}^K\sum_{i\neq j}
    \frac{|\omega_{i,j}^{(k)}|}{v_{z_{i,j}}}$\\$
    +
    \lambda_3\sum_{k<k'}\sum_{i\neq j}
    {v_{(w_{i,j}z_{i,j})}}^{-1}
    |\omega_{i,j}^{(k)}-\omega_{i,j}^{(k')}|$\footnote{ $z_{i,j}$, $w_{i,j}$ are binary variables for $i\neq j$ drawn independently from a Bernoulli distribution and $v_0$, $v_1$ are two constants such that $v_1>v_0>0$.}
    }
    }\\
    &&&&\\
    &&&&\\\cline{3-5}
    &&BJEMGM~\citep{gan2019bayesian} -- Section~\ref{ssec:coarse_bayesian}
&$\omega_{i,j}^{(k)}$&\multirow{3}{*}{
\shortstack[l]{
$\sum_{i=1}^p\sum_{k=1}^K \lambda_3\omega_{i,i}^{(k)}$\\$ + \sum_{i<j}  \log\big\{
\prod_{k=1}^K\frac{\lambda}{2\lambda_1}\exp(-|\omega_{i,j}^{(k)}|/\lambda_1)$\\$
\quad\qquad\qquad+\prod_{k=1}^K\frac{1-\lambda}{2\lambda_2}\exp(-|\omega_{i,j}^{(k)}|/\lambda_2)\big\}
$
}
    }\\
    &&&&\\
    &&&&\\
    &&&&\\\cline{3-5}
\hline
\multirow{11}{*}{\shortstack[l]{Fine\\-grained}}&\multirow{5}{*}{\shortstack[l]{Penalized\\MLE}}
&LASICH~\citep{saegusa2016joint} -- Section~\ref{ssec:fine_penalty}
&$\omega_{i,j}^{(k)}$&\multirow{2}{*}{\shortstack[l]{
$\lambda_1\sum_{k=1}^K\sum_{i\neq j}|\omega_{i,j}^{(k)}|$\\
$+\lambda_1\lambda_2\sum_{i\neq j}\{\sum_{k,k'}^KW_{k,k'}(\omega_{i,j}^{(k)}+\omega_{i,j}^{(k')})^2\}^{\frac{1}{2}}$
}}\\
&&&&\\\cline{3-5}
&&GEN-ISTA~\citep{price2021estimating} -- Section~\ref{ssec:fine_penalty}&
$\omega_{i,j}^{(k)}$&
\multirow{2}{*}{\shortstack[l]{
$\lambda_1\sum_{k=1}^K\sum_{i\neq j}|\omega_{i,j}^{(k)}|$\\
$+\lambda_2\sum_{q=1}^Q\frac{1}{|D_q|}\sum_{k,k'\in D_q}\|{\bm\Omega}^{(k)}-{\bm\Omega}^{(k)'}\|_F^2$
}}
\\
&&&&\\\cline{2-5}
&\multirow{2}{*}{\shortstack[l]{Neighborhood\\selection}}&
JSEM~\citep{ma2016joint} -- Section~\ref{ssec:fine_reg}
%JMMLE~\citep{majumdar2018joint}
&$\theta_{i,j}^{(k)}=-\omega_{i,j}^{(k)}/\omega_{i,i}^{(k)}$&$\sum_{j\neq i}\sum_{g\in\mathcal{G}_{i,j}}\lambda_{i,j}^{[g]}\|{\bm\theta}_{i,j}^{[g]}\|_2$\\
&&&&\\
\cline{2-5}
&\multirow{2}{*}{\shortstack[l]{Bayesian\\approach}}
&MRF~\citep{peterson2015bayesian} -- Section~\ref{sssec:LCS_bayesian}&$\omega_{i,j}^{(k)}$&$-\log\sbr{\prod_{k=1}^K |\Omega^{(k)}|^{(b-2)/2}\exp\{-2^{-1}\tr(\Omega^{(k)}{\bf D})\}}$
\\
%\cline{3-5}
%&&SLTs~\citep{oates2014joint} -- Section~\ref{sssec:LCS_bayesian}&$\omega_{i,j}^{(k)}$&
%\multirow{3}{*}{\shortstack[l]{
%$-\log\cbr{p(G^{(1)}|G^{0})\prod_{(k,k')\in E_T}p(G^{(k)}|G^{(k')})}$\\
%$p(G^{(k)}|G^{(k')})\propto\mathbbm{1}\{E^{(k)}\subseteq E^{(k')}\}\phi(G^{(k)})$\footnote{$\phi(G^{(k)})$ is a correction function that normalizes the prior.}
%}}\\
%&&&&\\
&&&&\\
\hline
\end{longtable}
\end{landscape}

\newpage

\section{Joint Graphical Models using Coarse-grained Structure}\label{section:coarse-grained_structure}

\begin{figure}
\centering
\begin{subfigure}{.5\textwidth}
  \centering
  \includegraphics[width=.9\linewidth]{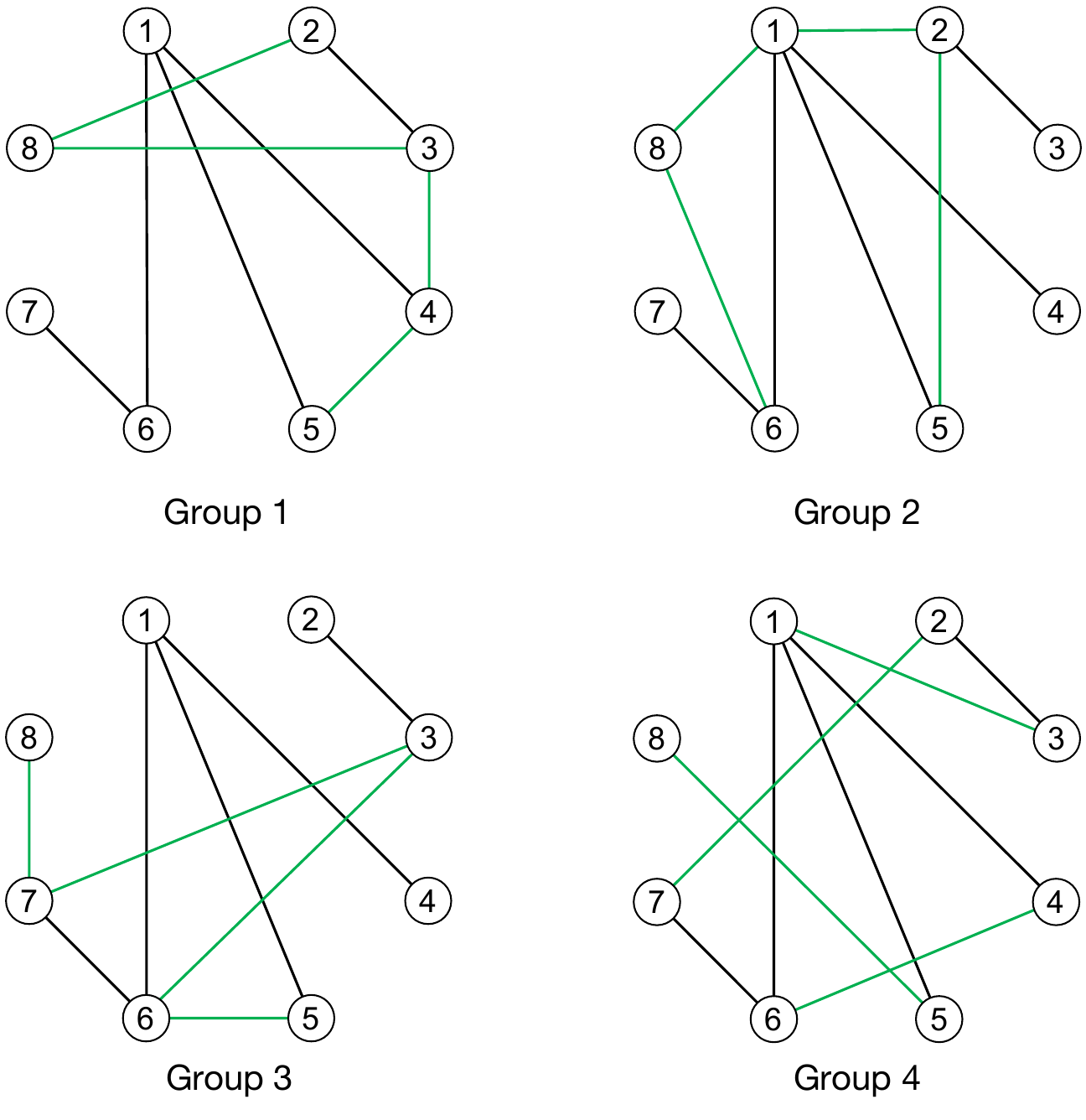}
  \label{fig:sub1}
  \caption{ }
\end{subfigure}%
\begin{subfigure}{.5\textwidth}
  \centering
  \includegraphics[width=.9\linewidth]{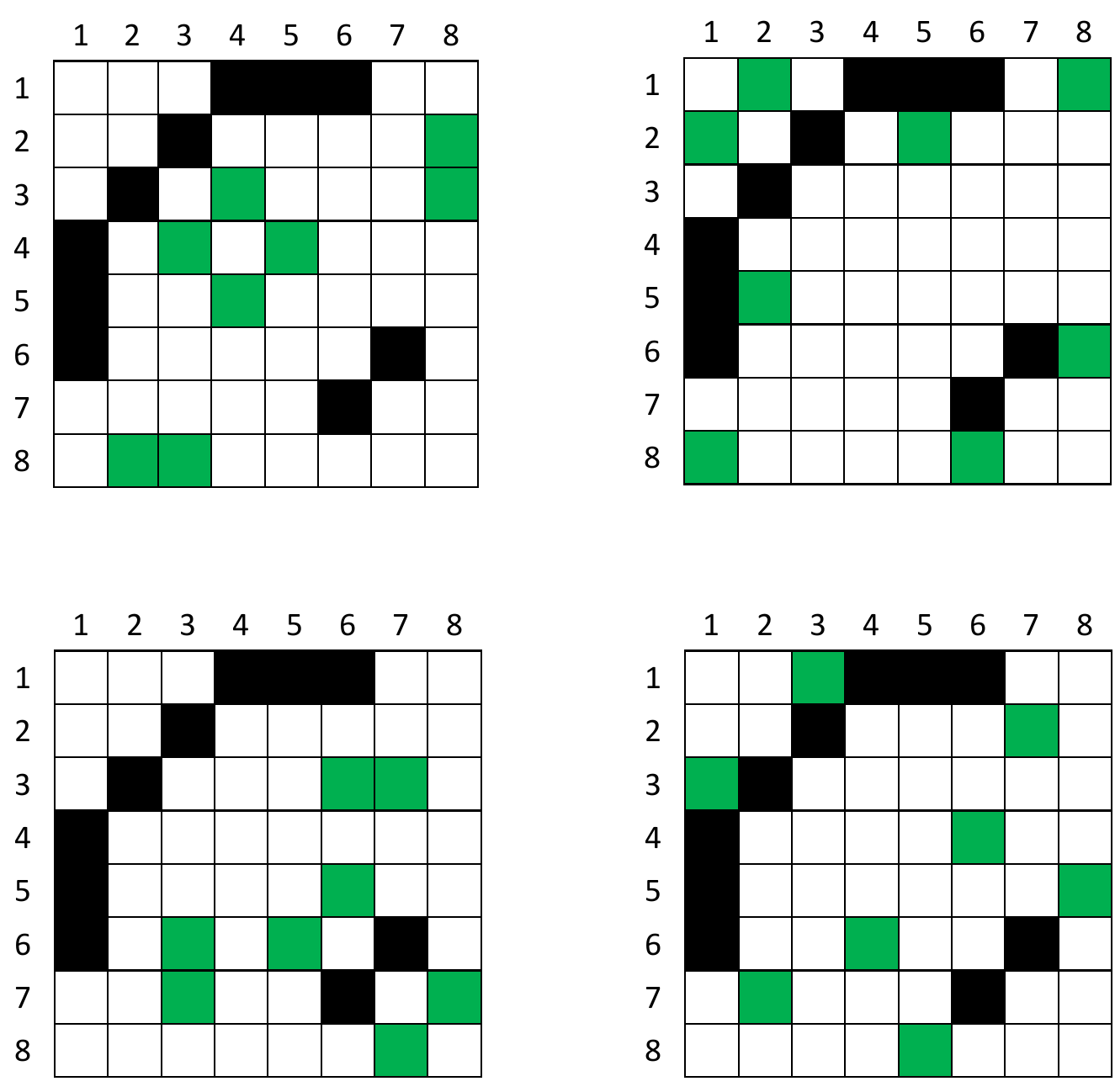}
  \label{fig:sub2}
  \caption{ }
\end{subfigure}
\caption{Graphical models with shared coarse-grained structure across groups. {\bf (a)} The black lines
  denote the common edges, while green lines denote individual edges. {Each graph has four unique individual edges that are not present in other graphs and four edges that are present in all graphs.}{\bf (b)} The corresponding adjacency matrices for each graph shown in (a).}
  \label{fig:globalstructure}
\end{figure}

We outline a variety of approaches for joint graphical model estimation that use prior knowledge of coarse-grained structures across groups. {As noted, for coarse-grained structures, all pairs of edge strengths are penalized/regularized in the same way, i.e., invariant to the group identity. The illustration of coarse-grained structure is shown in Figure~\ref{fig:globalstructure}. In contrast, fine-grained structure uses regularization/priors between edge strengths that vary across groups, e.g., using prespecified weights for pairs of groups. For comparison, the illustration of fine-grained structure is shown in Figure~\ref{fig:localstructure}.} The performance of the coarse-grained estimation procedure is improved using regularization that captures the common structure across the $K$ groups -- enabling the use of shared information across groups.  We will discuss two directions in detail: hierarchical regularizers and analogous Bayesian priors.

\subsection{Joint Graphical Models with Hierarchical Structure}\label{ssec:HSN}

\citet{guo2011joint} studied joint estimation of related precision matrices, where the precision matrices are assumed to be related through a hierarchical structure. Specifically, each entry in the precision matrix is the multiplication of a common component across $K$ groups and an individual component: $\omega_{i,j}^{(k)}=\theta_{i,j}\gamma_{i,j}^{(k)}$ for $i\neq j$ and $\omega_{i,i}^{(k)} = \gamma_{i,i}^{(k)}$, where $\theta_{i,j}$ is the shared component and $\gamma_{i,j}^{(k)}$ is the group-specific component. Thus, this approach enforces a common background structure. To encourage sparsity, an $\ell_1$-norm penalty term is also included as a regularizer, resulting in the following objective termed Joint Graphical Lasso (JGL):
\begin{align*}
    &\hat{{\bm\Theta}},\{\hat{\bm\Gamma}^{(k)}\}_{k=1}^K=\argmax\;{\ell}({\bm \Omega})-\lambda_1\sum_{i\neq j}\theta_{i,j}-\lambda_2\sum_{k=1}^K\sum_{i\neq j}|\gamma_{i,j}^{(k)}|,
\end{align*}
where $\lambda_1, \lambda_2$ are hyperparameters that control the scale of the penalty. Note that even when the common component $\theta_{i,j}$ is nonzero, an individual entry $\omega_{i,j}^{(k)}$ can still be set to zero by the $\ell_1$ penalty, which denotes a missing edge in the associated graph. {
It is worth pointing out that this method is non-convex and hence only convergence to local minima is guaranteed. \citet{danaher2014joint} introduced a similar method where the associated penalty functions are convex, which we discuss in Section~\ref{ssec:coarse_reg}. }

\citet{shan2020joint} proposed a Joint tWo-Level Graphical Lasso (JWLGL), which is a more expressive model that constructs two-level structures on both the set of common components and individual components. The algorithm further clusters the set of nodes $V$ into $M$ classes and imposes class specific structure: let $m$ and $m'$ be the classes to which nodes $i$ and $j$ belong, respectively. If $m\neq m'$, we have:
\begin{align*}
    \omega_{i,j}^{(k)}=\theta_{m,m'}^{(k)}\gamma_{i,j}^{m,m',(k)}; \qquad 
    \theta_{m,m'}^{(k)} = \alpha_{m,m'}\beta_{m,m'}^{(k)}.
\end{align*}
 If $m=m'$, we have $\theta_{m,m}^{(k)}=1$ and
\begin{align*}
    \omega_{i,j}^{(k)}=\theta_{m,m}^{(k)}\gamma_{i,j}^{m,m,(k)}; \qquad
    \gamma_{i,j}^{m,m,(k)} = \left\{\begin{array}{ll}
        \iota_{i,j}^{m,m}\rho_{i,j}^{m,m,(k)}  ,& i\neq j; \\
        1 ,& i=j.
    \end{array}\right.
\end{align*}
Without loss of generality, we assume $\alpha_{m,m'}\geq 0$ and $\iota_{i,j}^{m,m}\geq 0$ for $i\neq j$ and $m\neq m'$. Here, $\alpha_{m,m'}$ and $\iota_{i,j}^{m,m}$ denote the common components shared across $K$ groups, while $\beta_{m,m'}^{(k)}$ and $\rho_{i,j}^{m,m,(k)}$ denote the individual components that vary across groups.

\citet{lee2015joint} proposed a Joint Estimator of Multiple Precision matrices (JEMP) under an assumption that precision matrices decompose into the sum of two components: ${\omega}^{(k)}_{i,j}={\theta}_{i,j}+{\gamma}_{i,j}^{(k)}$. In contrast to the maximum likelihood, the estimation procedure of JEMP is motivated by the CLIME estimator~\citep{cai2011constrained}, which estimates a single precision matrix by solving the following optimization problem:
\begin{align}
    \hat{{\bm\Omega}}^{(k)}=\argmin \|{\bm\Omega}^{(k)}\|_1\label{eq:CLIME}
    \qquad
    \text{subject to}
    \qquad
    |\hat{{\bm\Sigma}}^{(k)}{\bm\Omega}^{(k)}-{\bm I}|_\infty\leq\xi,
\end{align}
where $\xi$ is a tuning parameter. The CLIME estimator finds a sparse $\hat{\bm\Omega}^{(k)}$ while ensuring that $\hat{{\bm\Sigma}}^{(k)}\hat{\bm\Omega}^{(k)}$ is close to an identity matrix. JEMP can be seen as a generalization of CLIME to a multi-group setting as it solves the following optimization problem:
\begin{align*}
      &\hat{{\bm \Theta}},\{\hat{\bm \Gamma}^{(k)}\}_{k=1}^{K}=\argmin\|{\bm \Theta}\|_1+v\sum_{k=1}^K\|{\bm \Gamma}^{(k)}\|_1,\\
        &\text{subject to  }\left|\frac{1}{K}\sum_{k=1}^K\cbr{\hat{{\bm\Sigma}}^{(k)}\rbr{{\bm \Theta}+{\bm \Gamma}^{(k)}}-{\bm I}}\right|_{\infty}\leq\lambda_1,\;\ \left|\hat{{\bm\Sigma}}^{(k)}\rbr{{\bm \Theta}+{\bm \Gamma}^{(k)}}-{\bm I}\right|_{\infty}\leq\lambda_2,\;\ \sum_{k=1}^K {\bm \Gamma}^{(k)}=0,\nonumber
\end{align*}
where ${\bm \Theta}$ denotes the common structure, i.e., the mean of the precision matrices ${K}^{-1}\sum_{k=1}^{K}{\bm\Omega}^{(k)}$, and ${\bm \Gamma}^{(k)}$ denotes the individual residual components ${\bm\Omega}^{(k)}-{\bf \Theta}$. In the above optimization problem, the first constraint regularizes the average difference and the second constraint regularizes the individual difference. Thus, the first constraint imposes a common structure across groups.  The prespecified weight $v$ controls the degree of uniqueness of each group, while $\lambda_1, \lambda_2$ are hyperparameters that measure group average and individual estimation quality, respectively.

\subsection{Regularization Approaches for Modeling Joint Structure}\label{ssec:coarse_reg}

Regularization-based approaches~\citep{danaher2014joint, bilgrau2020targeted, hao2017simultaneous, shan2018joint} do not assume the form of the common structure and individual structure, but instead impose similarity constraints across groups. For example, the Fused Graphical Lasso (FGL) and the Group Graphical Lasso (GGL)~\citep{danaher2014joint} add convex penalty terms to the log-likelihood function to learn a common structure:
\begin{align}
    &\hat{{\bm\Omega}}=\argmax\;{\ell}({\bm\Omega})-{\bm P}({\bm\Omega});\notag\\
    &{\bm P}_{\text{FGL}}({\bm\Omega})=\lambda_1\sum_{k=1}^K\sum_{i\neq j}|\omega_{i,j}^{(k)}|+\lambda_2\sum_{k<k'}\sum_{i,j}|\omega_{i,j}^{(k)}-\omega_{i,j}^{(k')}|\label{FGL};\\
    &{\bm P}_{\text{GGL}}({\bm\Omega})=\lambda_1\sum_{k=1}^K\sum_{i\neq j}|\omega_{i,j}^{(k)}|+\lambda_2\sum_{i\neq j}\cbr{\sum_{k=1}^K\rbr{\omega_{i,j}^{(k)}}^2}^{\frac{1}{2}}\label{GGL}.
\end{align}
The first penalty term in both ${\bm P}_{\text{FGL}}$ and ${\bm P}_{\text{GGL}}$ encourages model sparsity. The second term in ${\bm P}_{\text{FGL}}$ encourages  groups to have shared edge values, while the ${\bm P}_{\text{GGL}}$ penalty tends to be less restrictive and only encourages a shared sparsity pattern. { 
In addition, an R-package `JGL' is provided that implements both $FGL$ and $GGL$~\citet{danaher2014joint}.} \\\par

\citet{hao2017simultaneous} proposed simultaneous clustering and estimation (SCAN) procedure that addresses the case when the heterogeneous data are missing group labels, e.g., when the groups are latent or unknown. SCAN partitions the unlabeled data into $K$ clusters and simultaneously imposes a homogeneous structure across groups using the GGL penalty. Given $n$ unlabeled observations ${\bm x}_i,\;i=1,\hdots,n$, with the density function $l({\bm x},{\bm\mu},{\bm\Omega})=\sum_{k=1}^K\pi_kl_k({\bm x}; {\bm\mu}^{(k)},{\bm\Omega}^{(k)})$, where $\pi_k$ is the probability that ${\bm x}_i$ belongs to the $k$-th group and $l_k({\bm x}; {\bm\mu}^{(k)},{\bm\Omega}^{(k)})=\log\{\det({\bm\Omega}^{(k)})\}-\tr\{({\bm x}-{\bm\mu}^{(k)})({\bm x}-{\bm\mu}^{(k)})^\top{\bm\Omega}^{(k)}\}$. SCAN solves the following optimization problem:
\begin{align*}
    &\{\hat{\pi}_k\}_{k=1}^K,\hat{\bm\mu},\hat{\bm\Omega}=\argmax \frac{1}{n}\sum_{i=1}^n\log\left(l({\bm x}_i,{\bm\mu},{\bm\Omega})\right)-{\bm P}_{SCAN}({\bm\mu},{\bm\Omega});\\
    &{\bm P}_{SCAN}({\bm\mu},{\bm\Omega})=\lambda_1\sum_{k=1}^K\sum_{i\neq j}|\omega_{i,j}^{(k)}|+\lambda_2\sum_{i\neq j}\left\{\sum_{k=1}^K\rbr{\omega_{i,j}^{(k)}}^2\right\}^{\frac{1}{2}}+\lambda_3\sum_{k=1}\sum_{i=1}^p|\mu_i^{(k)}|.
\end{align*}
{
Note that the first two terms of ${\bm P}_{SCAN}({\bm\mu},{\bm\Omega})$ correspond to the GGL penalty function and the third term is the $\ell_1$-norm penalty, used for encouraging the sparsity of the mean vectors. Such regularization is common in the context of high-dimensional data, where many variables contain limited information about the clustering structure. Hence, placing a sparse penalty function realizes selection of informative variables~\citep{pan2007penalized, sun2012regularized}.}

There are several other methods that use the regularization approach to estimate the joint structure. \citet{shan2018joint} proposed the Joint Adaptive Graphical Lasso (JAGL) procedure that introduces a weighted $\ell_1$ penalty term to tackle problems with unbalanced data. \citet{bilgrau2020targeted} proposed Targeted Fused Ridge Estimator (TFRE) that uses an additional $\ell_2$ penalty term that incorporates target matrices as prior information to stabilize the estimation process. {
In addition, an R-package `rags2ridges' provides an implementation of TFRE~\citep{bilgrau2020targeted}.}
%{Finally, the joint greedy equivalence search (JointGES)
%procedure
%uses an $\ell_0$ penalty term to jointly estimate directed acylic graphs~\citep{wang2020high}.}
\\\par

\subsection{Bayesian Methods for Joint Gaussian Graphical Model Estimation}\label{ssec:coarse_bayesian}

We briefly overview Bayesian methods for joint Gaussian graphical model estimation. Bayesian formulations of graphical models use priors to encourage desired properties for model selection. For example, the spike-and-slab prior is commonly used in practice to encourage sparsity in precision matrices. In general, the probabilistic counterpart to the penalty function follows the relation $p({\bm \Omega})\propto \exp(-{\bm P}({\bm\Omega}))$, where $p({\bm \Omega})$ is the prior distribution of the precision matrix and ${\bm P}({\bm\Omega})$ denotes the penalty function. By the Bayes' rule, the posterior distribution is proportional to the product of the likelihood and prior distribution. Therefore, finding a maximum a posteriori probability (MAP) estimate is equivalent to obtaining the estimate by maximizing the log-likelihood (MLE) with an additional penalty function. Several works~\citep{tan2017bayesian,li2019bayesian,yajima2014detecting,mitra2016bayesian,peterson2015bayesian,lin2017joint} have addressed Bayesian graphical model estimation by designing priors that incorporate structural information.
In the Gaussian graphical model case, a Wishart prior~\citep{roverato2002hyper,atay2005monte,lenkoski2011computational, mohammadi2015bayesian} is often placed on the precision matrix. This prior is a conjugate prior for the Gaussian likelihood, i.e., for a Gaussian likelihood, the posterior distribution remains Wishart. Furthermore, the normalizing constant of the posterior distribution has an explicit form when the graph $G$ is decomposable, that is, when the index set $V$ of a graph can be partitioned into three disjoint nonempty sets  $V=A\cup S\cup B$ and (i) $S$ is a clique,  (ii) $S$ separates $A$ and $B$, (iii) $A\cup S$ and $S\cup B$ form decomposable subgraphs.\\\par

\citet{li2019bayesian} proposed the joint spike-and-slab graphical lasso prior, designed to encourage global sparse structure. In a related work, \citet{tan2017bayesian} placed a multiplicative prior on the adjacency matrices,  where the distribution of each edge depends on the multiplication of the values of two end nodes. This prior not only encourages sparsity, but also allows users to specify the degree of connections. We discuss local common structure methods~\citep{peterson2015bayesian,lin2017joint} in Section~\ref{sssec:LCS_bayesian} and the differential graph methods~\citep{yajima2014detecting,mitra2016bayesian} in Section~\ref{ssec:diffnetwork_reg}.  We will not go into details about the Bayesian formulation of graphical models, but instead give a high-level overview of various approaches. \\\par

Consider a single group setting with ${\bm\Omega}=\{{\bm\Omega}^{(1)}\}$. From a Bayesian perspective, the lasso regularizer can be viewed as a Laplace prior~\citep{marlin2009sparse,wang2012bayesian} and is formulated as:
\begin{equation}\label{eq:laplace_prior}
p({\bm \Omega}\mid\lambda)
\propto\prod_{i\neq j}
\frac{\lambda}{2}\exp\left(-\lambda|\omega_{i,j}|\right)
\prod_{i=1}^p
\left\{
    \frac{\lambda}{2}
    \exp\left(-\frac{\lambda}{2}\omega_{i,i}\right)
    \mathbbm{1}_{(\omega_{i,i}>0)}
\right\}
\mathbbm{1}({\bm \Omega}\succ 0),
\end{equation}
where  $\mathbbm{1}({\bm \Omega}\succ 0)$ restricts the precision matrix to be positive definite. The term $\mathbbm{1}_{(\omega_{i,i}>0)}$ ensures that the diagonal entries are non-negative and hence preserves the positive definiteness of ${\bm \Omega}$. When taking the logarithm $\log p({\bm \Omega}\mid\lambda)$, the first product is equal to the lasso regularizer. Therefore, when computing  the MAP estimate, the logarithm of the Laplace prior along with the log-likelihood is the penalized MLE estimator with lasso penalty function. In the multiple group case, to promote the group similarity between the precision matrices, \citet{li2019bayesian} converted the GGL and FGL penalties to structural priors. 

In Bayesian inference, other than computing the MAP estimator, we are also interested in the posterior mean, mode, and samples. In this case, the shrinkage priors are not enough to produce sparse posterior samples (or mean and mode) because the posterior does not concentrate on sparse parameters. Therefore, additional thresholding is required to obtain sparsity. As an alternative, one may use the spike-and-slab prior~\citep{mitchell1988bayesian} to promote the sparsity pattern in the posterior. Consider a single group ${\bm\Omega}=\{{\bm\Omega}^{(1)}\}$, the spike-and-slab prior is a hierarchical mixture prior formulated as:
\begin{align}\label{eq:spike-slab}
&p({\bf z}\mid\lambda) = \prod_{i\neq j}\Ber(z_{i,j}\mid \lambda);\\
&p({\bm \Omega}\mid{\bf z}) 
= 
\prod_{i\neq j}
(1-z_{i,j})\delta(\omega_{i,j}) + z_{i,j}\mathcal{N}(\omega_{i,j}\mid 0,\sigma^2),\notag
\end{align}
where $\delta(\cdot)$ denotes the delta function. If $z_{i,j}=0$, we restrict the variable to be zero. One may also replace the delta function with a normal distribution with small variance, which approximates the delta function. In the multiple group case, a set of latent indicators following the spike-and-slab distribution adaptively control the value of the FGL (resp., GGL) penalty~\citep{li2019bayesian}, namely the Doubly Spike-and-Slab Joint Graphical Lasso (DSS-JGL). Consider two constants $v_1>v_0>0$ and $z_{i,j}$, $w_{i,j}$ are binary variables for $i\neq j$. We assume that each $z_{i,j}$ and $w_{i,j}$ are drawn independently from a Bernoulli distribution with a specific parameter. The DSS-JGL prior is represented as:
\begin{equation}\label{DSSJGL}
    -\log p({\bm\Omega}\mid{\bf z},{\bf w}) \propto
    \lambda_1 \sum_{k=1}^K\sum_{i=1}^p
    |\omega_{i,i}^{(k)}|
    +
    \lambda_2\sum_{k=1}^K\sum_{i\neq j}
    \frac{|\omega_{i,j}^{(k)}|}{v_{z_{i,j}}}
    +
    \lambda_3\sum_{k<k'}\sum_{i\neq j}
    \frac{1}{v_{(w_{i,j}z_{i,j})}}|\omega_{i,j}^{(k)}
    -
    \omega_{i,j}^{(k')}|,
\end{equation}
where the third term can also be replaced by the group lasso penalty function, similar to~\eqref{GGL}. We can choose $v_0$ to be small, so that when $z_{i,j}=0$ for $i\neq j$, the second term in~\eqref{DSSJGL} will be large, forcing the posterior to be zero. Similar behavior also follows for the joint regularization term in~\eqref{DSSJGL} when either $z_{i,j}$ or $w_{i,j}$ is zero. {Additionally, an R-package `SSJGL' provides an implementation of DSS-JGL~\citep{li2019bayesian}.} \\\par

{Although the Bayesian approaches introduced above provide expressive structures for joint estimation of multiple graphical models, theoretical guarantees that characterize convergence rates are lacking in general. \citet{gan2019bayesian} provided guarantees on the structure recovery and the convergence rate in $\ell_\infty$ norm. Specifically, they proposed Bayesian Joint Estimation of Multiple Graphical Models (BJEMGM) that extends the spike-and-slab prior to multiple graphs, but in a different setting compared to~\citet{li2019bayesian}. Let $z_{i,j}$ be i.i.d.~samples drawn from $\Ber(\lambda)$ with $0\leq\lambda\leq1$. The prior on ${\bf \omega}_{i,j}=\{\omega_{i,j}^{(1)},\ldots,\omega_{i,j}^{(K)}\}$ is defined as
\begin{align}\label{eq:BJEMGM}
    -\log p({\bf \omega}_{i,j}\mid z_{i,j},\lambda_1,\lambda_2)\propto-\log\cbr{\prod_{k=1}^Kz_{i,j}\frac{\lambda_1}{2}\exp\left(-\lambda_1|\omega_{i,j}^{(k)}|\right)+\prod_{k=1}^K(1-z_{i,j})\frac{\lambda_2}{2}\exp\left(-\lambda_2|\omega_{i,j}^{(k)}|\right)}.
\end{align}
The prior on the diagonal entries $\omega_{i,i}^{(k)}$, $i=1,\ldots,p$, $k=1,\ldots,K$, is the same as the second term of~\eqref{eq:laplace_prior}
with parameter $\lambda$ replaced by $\lambda_3$. Marginalizing over ${\bf z}$, the log of the prior distribution is expressed as
\begin{align*}
-\log p({\bm\Omega}\mid\lambda,\lambda_1,\lambda_2,\lambda_3) = \sum_{i=1}^p\sum_{k=1}^K \lambda_3\omega_{i,i}^{(k)} + \sum_{i<j}  \log\left(
\prod_{k=1}^K\frac{\lambda}{2\lambda_1}\exp(-|\omega_{i,j}^{(k)}|/\lambda_1)+
\prod_{k=1}^K\frac{1-\lambda}{2\lambda_2}\exp(-|\omega_{i,j}^{(k)}|/\lambda_2)
\right).
\end{align*}
From the modeling perspective, the prior in~\eqref{DSSJGL} additionally enforces the similarity of inverse covariance values $\omega_{i,j}^{(k)}$ for $k=1,\ldots,K$ and $i\neq j$, while ~\eqref{eq:BJEMGM} only constructed a shared latent Bernoulli variable $z_{i,j}$ across $K$ groups that controls the sparsity of $\omega_{i,j}^{(k)}$ for $k=1,\ldots,K$.}

\section{Joint Graphical Models using Fine-grained Structure}
\label{section:fine-grained_structure}

\begin{figure}
\centering
\begin{subfigure}{.5\textwidth}
  \centering
  \includegraphics[width=.9\linewidth]{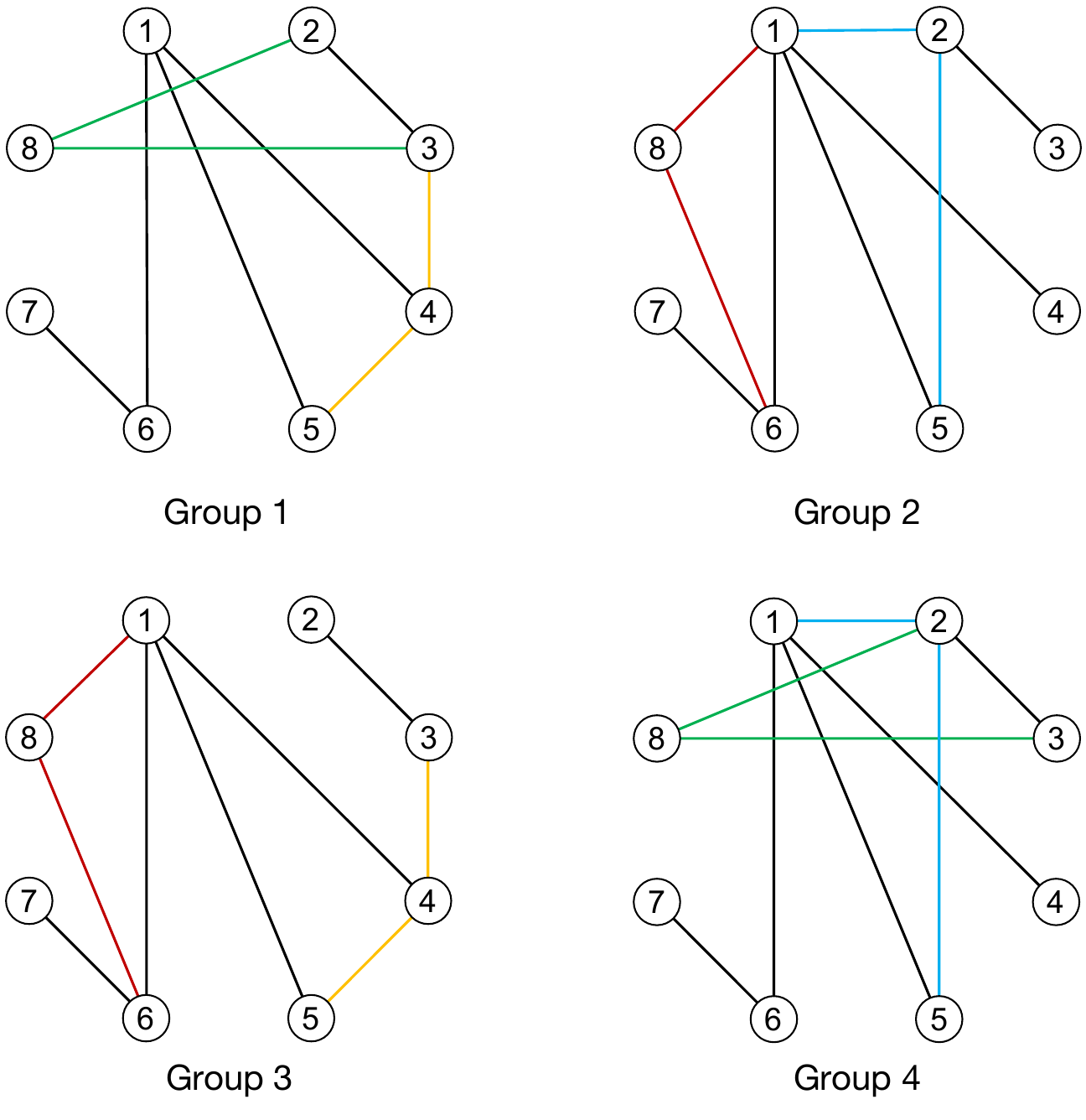}
  \label{fig:sub3}
  \caption{}
\end{subfigure}%
\begin{subfigure}{.5\textwidth}
  \centering
  \includegraphics[width=.9\linewidth]{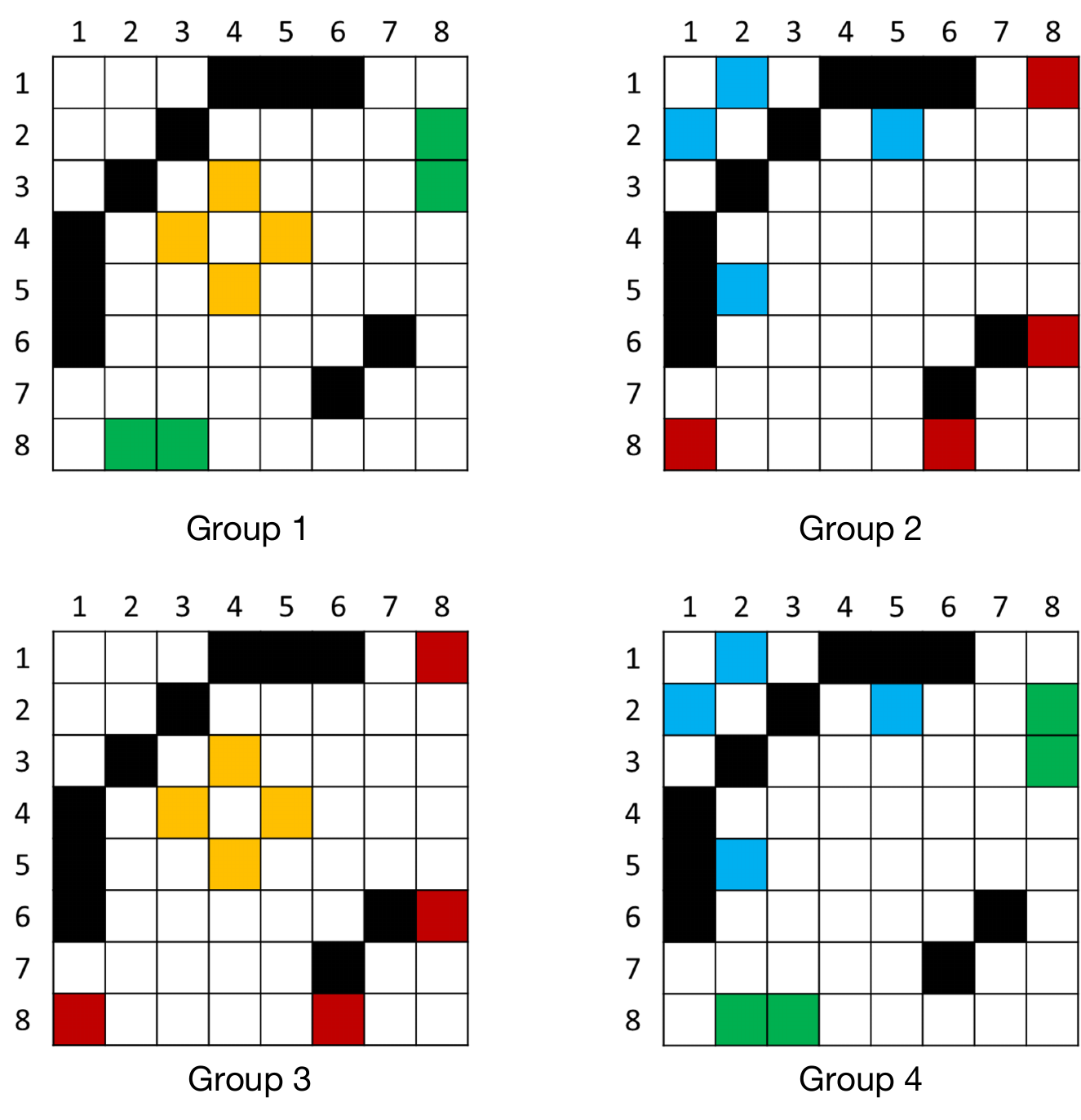}
  \label{fig:sub4}
  \caption{}
\end{subfigure}
\caption{Graphical models with shared fine-grained structure across groups. {\bf (a)} The black lines denote the common edges. Green colored lines represent the common structure of Group 1 and Group 4, yellow lines of Group 1 and Group 3, red lines of Group 2 and Group 3, and blue lines of Group 2 and Group 4. {\bf (b)} The corresponding adjacency matrices for each graph shown in (a), using the same colors for groups of shared edges.}
\label{fig:localstructure}
\end{figure}

When estimating coarse-grained joint graphical models, all pairs of edge strengths are penalized equally; the relationship between $\omega_{i,j}^{(k)}$ and $\omega_{i,j}^{(k')}$ and the relationship between $\omega_{i,j}^{(k)}$ and $\omega_{i,j}^{(k'')}$, $k\neq k'$, $k'\neq k''$ are assumed to be equal. However, in many real world settings, some subsets of groups share a local structure that does not appear across all groups. Figure~\ref{fig:localstructure} visualizes the adjacency matrices related to edge sets ${E}^{k},\;k=1,\hdots,4$, of graphical models that are not globally similar, but with subsets of groups that share a local structure. The most common approach in modeling such graphical models is to incorporate this prior knowledge of the relation between subgroups into the estimators~\citep{ma2016joint,saegusa2016joint}. We briefly outline some approaches for joint graphical model estimation with fine-grained shared structure.

\subsection{Entry-wise Structural Information}\label{ssec:fine_reg}

Given the relation information $\mathcal{G}=\cup_{1\leq i<j\leq p}\mathcal{G}_{i,j}$, where $\mathcal{G}_{i,j}$ is a set that encodes the group relations of node $i$ and node $j$, Figure~\ref{fig:localstructure} (a) illustrates an example of entry-wise structural information. Consider the pair of nodes $(i,j)=(3,4)$ in Figure~\ref{fig:localstructure} (a): both graphs of Group 1 and Group 2 have edges (yellow lines) connecting these two nodes, while Group 3 and Group 4 do not have an edge in between. Then the corresponding relation information $\mathcal{G}_{3,4}=\mathcal{G}_{4,3}$ is $\{\{1,3\},\{2,4\}\}$. Similarly, for the pair of nodes $(i,j)=(2,8)$, the graphs of Group 1 and Group 4 have an edge (green line) connecting $(2,8)$, while graphs of Group 2 and Group 3 do not have edge between node $(2,8)$. Then, the corresponding relation information $\mathcal{G}_{2,8} = \mathcal{G}_{8,2}$ is $\{\{1,4\},\{2,3\}\}$. The relation information of graphs in Figure~\ref{fig:localstructure} is $\mathcal{G}_{i,j}=\{\{1,3\},\{2,4\}\}$ for $(i,j)\in\{(3,4),(4,3),(4,5),(5,4),(1,2),(2,1),(2,5),(5,2)\}$;  $\mathcal{G}_{i,j}=\{\{1,4\},\{2,3\}\}$ for $(i,j)\in\{(2,8),(8,2),(3,8),(8,3), (1,8), (8,1), (6,8), (8,6)\}$; and $\mathcal{G}_{i,j}=\{\{1,2,3,4\}\}$ for the rest of the node pairs.\\\par

\citet{ma2016joint} proposed a joint structural estimation method (JSEM) to estimate edge sets $\{\hat{E}^{(k)}\}_{k=1}^K$ by modifying the neighborhood selection algorithm~\citep{meinshausen2006high} to incorporate structural information. Once the edge sets are estimated, each precision matrix is estimated by maximizing the group specific likelihood constrained to have zeros indexed by $\{\hat{E}^{(k)}\}_{k=1}^K$. We will briefly introduce the neighborhood selection method and then show how JSEM extends it to multiple graphical models. \\\par

The neighborhood selection algorithm estimates the conditional independence structure, which is encoded in the pattern of zeros of the precision matrix under a Gaussian model, by solving a collection of regression problems. See Chapter 12 in~\citet{maathuis2018handbook} and references therein. Suppose the $p$-dimensional random variable ${\bm x}$ follows a normal distribution $\mathcal{N}({\bm\mu},({\bm\Omega})^{-1})$ with an associated graph $G=(V,E)$.  In this case, we can express $x_i$, $i\in V$, as a linear function of other nodes:
\[
x_i = \sum_{j\in V\text{\textbackslash}\{i\}}\theta_{ij}x_j + \varepsilon_j
,
\]
where $\varepsilon_j$ is independent of $x_j$, $j\in V\text{\textbackslash}\{i\}$, if and only if $\theta_{ij}=-\omega_{i,j}/\omega_{i,i}$. Therefore, the optimal prediction of ${x}_i$ given the remaining variables can be formulated as the following optimization problem:
\begin{align}\label{NeiSel}
    &\{\hat{\theta}_{i,j}\}_{j\in V\text{\textbackslash}\{i\}} =\argmin \;\mathbb{E}\left({ x}_i-\sum_{j\in V\text{\textbackslash}\{i\}}\theta_{i,j}{ x}_j\right)^2.
\end{align}
Since the distribution of ${\bm x}$ is not known, the expectation term in~\eqref{NeiSel} can not be computed. Let ${\bm X}$ be a $n\times p$ matrix that collects $n$ i.i.d.~observations of ${\bm x}$. The $\ell_1$  penalized empirical optimization objective is given as:
\begin{align}\label{NS_empirical}
    &\hat{\bm{\Theta}}_i=\argmin_{{\bm\Theta}_i:\;\theta_{i,i}=0}\frac{1}{n}\|{\bm X}_i-{\bm X}{\bm\Theta}_i\|_2^2+\lambda\|{\bm\Theta}_i\|_1,
\end{align}
where ${\bm \Theta}_i$ is a $p$-dimensional vector ${\bm \Theta}_i=(\theta_{i,1},\hdots,\theta_{i,p})^\top$, $\theta_{i,i}=0$, and ${\bm X}_i$ is the $i$-th column of the matrix ${\bm X}$. To estimate multiple models, JSEM extends the neighborhood selection approach in \eqref{NS_empirical} by solving the following optimization problem:
\begin{align}
    &\hat{\bm\Theta}=\argmin_{{\bm\Theta}_i^{(k)}:\theta_{i,i}^{(k)}=0,\;k=1,\hdots,K} \sum_{k=1}^K\frac{1}{n_k}\|{\bm X}_i^{(k)}-{\bm X}^{(k)}{\bm \Theta}_i^{(k)}\|_2^2+2{\bm P}_{JSEM}(\{{\bm\Theta}_i^{(k)}\}_{k=1}^K)\label{JSEM};\\
    &{\bm P}_{JSEM}(\{{\bm\Theta}_i^{(k)}\}_{k=1}^K)=\sum_{j\neq i}\sum_{g\in\mathcal{G}_{i,j}}\lambda_{i,j}^{[g]}\|{\bm\theta}_{i,j}^{[g]}\|_2\notag,
\end{align}
where ${\bm\Theta}=\{{\bm\Theta}^{(1)},\hdots,{\bm\Theta}^{(K)}\}$, the penalty term incorporates the relation information $\mathcal{G}$, and $\lambda_{i,j}^{[g]}$ is the group-specific tuning parameter. The penalty function ${\bm P}_{JSEM}$ resembles the GGL penalty in \eqref{GGL}, except that the norm is now placed on a subset of groups provided by the relation information $\mathcal{G}$. Minimizing \eqref{JSEM} results in the following estimated edge sets $$\hat{E}^{(k)}=\{(i,j):1\leq i <j\leq p,\hat{\theta}_{i,j}^{(k)}\neq 0 \text{ or }\hat{\theta}_{j,i}^{(k)}\neq0\},\quad k=1,\hdots,K.$$
Given $\hat{E}^{(k)}$, we define $\mathcal{S}^+_{\hat{E}^{(k)}}=\{{\bm\Omega}:{\bm\Omega} \succ 0,\; \omega_{i,j}=0,\; \forall (i,j)\not\in \hat{E}^{(k)}\text{ and } i\neq j\}$. The precision matrix set ${\bm\Omega}$ is finally estimated by maximizing the log-likelihood with constraints that enforce the sparsity pattern:
\begin{align}\label{JSEM_MLE}
    &\hat{\bm\Omega}^{(k)}=\argmax_{{\bm\Omega}^{(k)}\in\mathcal{S}^+_{\hat{E}^{(k)}}}\log\{\det({\bm\Omega}^{(k)})\}-\tr(\hat{\bm\Sigma}^{(k)}{\bm\Omega}^{(k)}),\quad k=1,\hdots,K.
\end{align}
Note that we can apply JSEM only when element-wise structural relation information is given. However, when not all element-wise structural relation information is provided, one can still use the GGL penalty on subsets of groups for which prior information is available.

\subsection{Group-wise Structural Information}
\label{ssec:fine_penalty}

As obtaining entry-wise structural information is sometimes challenging, another approach is to use the relationship between groups, described by weights, in estimation. For example, suppose that $K=3$ and we have the following relationships between groups: Group $1$ and Group $2$ share similarity, Group $1$ and Group $3$ also share similarity, but Group $2$ and Group $3$ are unlikely to be similar. To this end, \citet{saegusa2016joint}  proposed LAplacian Shrinkage for Inverse Covariance matrices from Heterogeneous populations (LASICH), that uses a weighted graph $G_g=(\Gamma,E,W)$ to record the relations between groups. The node set $\Gamma$ denotes $K$ groups, the edge set $E$ captures the relations between groups, and the weight set $W:E\rightarrow\mathbb{R}_+$ represents the similarities between groups. Then, a Laplacian penalty function is placed on the objective function to impose group similarity. The optimization problem is formulated as follows:
\begin{align*}
    &\hat{\bm\Omega}=\argmax\;{\ell}({\bm\Omega})-{\bm P}_{LASICH}({\bm\Omega}),\\
    &\text{subject to }{\bm\Omega}^{(k)}=({\bm\Omega}^{(k)})^\top,\;{\bm\Omega}^{(k)}\in\mathcal{S}\succ 0,\quad k=1,\hdots,K\nonumber;\\
    &{\bm P}_{LASICH}({\bm\Omega})=\lambda_1\sum_{k=1}^K\sum_{i\neq j}|\omega_{i,j}^{(k)}|+\lambda_1\lambda_2\sum_{i\neq j}\left\{\sum_{k,k'}^KW_{k,k'}\left(\omega_{i,j}^{(k)}+\omega_{i,j}^{(k')}\right)^2\right\}^{\frac{1}{2}},
\end{align*}
where the first penalty term in ${\bm P}_{LASICH}({\bm\Omega})$ promotes the sparsity pattern and the second term encourages the similarities within subsets of groups. { 
In the case where the weight set is not available, \citet{saegusa2016joint} further proposed a two-stage algorithm, called Hierarchical Clustering LAISCH (HC-LAISCH),  that first uses hierarchical clustering to learn relations between groups and then applies LAISCH. Furthermore, under that Gaussian assumption, the estimates from HC-LAISCH and LAISCH share the same consistency properties.
} 
Although the approaches in \citet{ma2016joint}, and \citet{saegusa2016joint} require prior information on the group relations, or the prior information is obtained by another algorithm, they provide a more flexible structure than the global structure methodologies, such as GGL~\citep{danaher2014joint} and \citet{guo2011joint}. In particular, the global structure approach can be viewed as a special case of a local structure approach with homogeneous structural information. 

While the work introduced earlier required prior information about the group relations, which may not be available for most cases, Graphical Elastic Net Iterative Shrinkage Thresholding Algorithm (GEN-ISTA)~\citep{price2021estimating} jointly estimates graphs and group membership via k-means clustering. That is, GEN-ISTA  further clusters groups $k=1,\ldots,K$ into $Q$ classes. Let $D_q$, $q=1,\ldots,Q$, be the sets that contain group memberships. The objective function for GEN-ISTA is
\begin{align}
    &\argmax_{{\bm\Omega}, {\bm D}} {\ell}({\bm\Omega}) -{\bf P}_{GEN-ISTA}({\bm\Omega},{\bm D});\label{eq:gen}\\
    &{\bf P}_{GEN-ISTA}({\bm\Omega},{\bm D})=\lambda_1\sum_{k=1}^K\sum_{i\neq j}|\omega_{i,j}^{(k)}|+\lambda_2\sum_{q=1}^Q\frac{1}{|D_q|}\sum_{k,k'\in D_q}\|{\bm\Omega}^{(k)}-{\bm\Omega}^{(k)'}\|_F^2.\notag
\end{align}
It is easy to see that when we fix ${\bm\Omega}$ and optimize ${\bm D}$, then~\eqref{eq:gen} becomes a k-means clustering objective. {
In contrast, when we fix ${\bm D}$ and optimize ${\bm \Omega}$ alone, the problem~\eqref{eq:gen} reduces to a joint graphical model with a variant of FGL~\citep{danaher2014joint}. A linear rate of convergence can be shown for the algorithm that minimizes~\eqref{eq:gen} by alternating minimization over  ${\bm D}$ and ${\bm \Omega}$~\citep{price2021estimating}.}

\subsection{Bayesian Approach}\label{sssec:LCS_bayesian}

We introduce two Bayesian methods that construct priors to constrain the similarities within subsets of groups. {
Both approaches do not require prior information to build group relations. This property is particularly desirable because, in most cases, we may not have the structural information for $K$ groups of data. }\\\par
\citet{peterson2015bayesian} proposed the Markov Random Field (MRF) prior for the graphs $G^{(k)}=(V,E^{(k)}),k=1,\hdots,K$ to encourage the selection of edge indices in related graphs. In addition, the selection of edge indices is controlled by two variables: a random matrix ${\bm \Theta}\in\mathbb{R}^{K\times K}$, whose $k,k'$-th entry measures the degree of similarity between group $k$ and $k'$ and an edge-specific coefficient vector ${\bm v}$ reflecting the probability of the corresponding edge being selected. 
{ 
Let ${\bf e}_{i,j}\in\mathbb{R}^K$, $1\leq i<j\leq p$, be a binary vector indicating the existence of an edge between node $i$ and $j$ among $K$ groups.} The prior for ${\bf e}_{i,j}$ is expressed as
\begin{align*}
p({\bf e}_{i,j}\mid v_{i,j},{\bm\Theta})\propto\exp(v_{i,j}{\bm 1}^T{\bf e}_{i,j}+{\bf e}_{i,j}^T{\bm \Theta}{\bf e}_{i,j}),\qquad 1\leq i<j\leq p,
\end{align*}
where $v_{i,j}$ controls the probability that an edge between the $i$-th and $j$-th nodes is selected. Specifically, ${\bm v}$ controls the sparsity of graphs. { 
The joint prior for the graphs ${\bf G}=\{G^{(1)},\ldots G^{(K)}\}$ is}
\[
p({\bf G}\mid{\bf v},{\bm\Theta}) = \prod_{i<j}p({\bf e}_{i,j}\mid v_{i,j},{\bm\Theta}).
\]
{The Beta prior is placed on the elements of ${\bm v}$. Specifically, we have that $v_{i,j} \sim \Beta(1,4)$, which results in a sparse graph.
Meanwhile, the spike-and-slab prior is placed on the group similarity matrix $\Theta$, conditioned on the latent random variables ${\bm z}=(z_{i,j})_{i<j}$,  to allow discrimination between zero similarity and positive similarity, where $z_{i,j}$, $1\leq i<j\leq K$, is a binary random variable denoting the relation between groups $i$ and $j$. If $ z_{i,j}=1$, the two groups are related, otherwise they are not. Then, the prior on ${\bm\Theta}$ is defined as
\begin{align*}
&p({\bm\Theta}\mid{\bm z})=\prod_{i<j}p(\theta_{i,j}\mid z_{i,j});\\
&p(\theta_{i,j}\mid z_{i,j})=
(1- z_{i,j})\delta +  z_{i,j}
\frac{\beta}{\Gamma(\alpha)}\theta_{i,j}^{\alpha-1}\exp(-\beta\theta_{i,j}),
\end{align*}
where $\Gamma(\cdot)$ is the Gamma function and $\alpha,\beta$ are hyperparameters. Noting that the distribution of ${\bm z}$ determines the relatedness between groups, \citet{peterson2015bayesian} constructed a Bernoulli prior on ${\bm z}$:
 \begin{align*}
     &p({\bm z}\mid w)=\prod p( z_{i,j}\mid w);\\
     &p( z_{i,j}\mid w) = 
     w^{ z_{i,j}}(1-w)^{1- z_{i,j}},
 \end{align*}
where $w\in[0,1]$ is a hyperparameter. \citet{shaddox2020bayesian} recently proposed an alternative prior on $\bm z$ in the setting where data types are different.}
Finally, we apply the G-Wishart prior distribution to the inverse covariance matrices conditional on the graph structures $G^{(k)}=(V,E^{(k)}),\;k=1,\hdots,K$:
{
\[
p(\bm{\Omega}\mid{\bm G}, b, {\bf D}) \propto \prod_{k=1}^K |\Omega^{(k)}|^{(b-2)/2}\exp\left\{-2^{-1}\tr(\Omega^{(k)}{\bf D})\right\}, \qquad \Omega^{(k)}\in P_{G^{(k)}},
\]
where ${\bf D}$ is a preselected positive definite matrix and $b>2$ is a constant. The set $P_{G^{(k)}}$ contains all positive definite matrices that have the edge structure $E^{(k)}$. \citet{shaddox2018bayesian} proposed a similar framework as~\citet{peterson2015bayesian}, but adopted a continuous shrinkage prior, instead of the G-Wishart prior used in~\citet{peterson2015bayesian}, resulting in a computationally more efficient procedure.}\\\par

{
\citet{lin2017joint} applied the MRF prior to jointly estimate multiple graphical models but in slightly different setting -- they adopted a Bayesian version of the neighborhood regression~\citep{meinshausen2006high}, see~\eqref{NeiSel}, and proposed a hierarchical prior on the set of regression coefficients ${\bm\Theta}=\{{\bm\Theta}^{(1)},\ldots,{\bm\Theta}^{(K)}\}$. To encourage the sparsity pattern, a spike-and-slab prior, see~\eqref{eq:spike-slab}, is placed on ${\bm\Theta}$ conditioning on binary latent factors ${\bm z}=\{{\bm z}_{i,j}\}$ with ${\bm z}=({ z}_{i,j}^{(1)},\ldots,{ z}_{i,j}^{(K)})\in\{0,1\}^K$, $i< j$. \citet{lin2017joint} studied the setting where the group index is a tuple $(s,t)$ with $s\in\mathcal{S} \subseteq \NN$ being the location and $t\in\mathcal{N} \subseteq \NN$ being the time. Intuitively, groups with the same time index $t$ should have similar graph structures, while groups with the same location $s$ and small pairwise time difference, i.e., $|t-t'|=1$ should have similar graph structures. Let $A_s=\{(s,t,s',t'):s\neq s', t=t'\}$ and $B_t=\{(s,t,s',t'):s=s', |t-t'|=1\}$. Let ${\bm\lambda}=\{\lambda_1,\lambda_2,\lambda_3\}$ be a set of hyperparameters.
The indicator function $\mathbbm{1}_a(x)$ outputs $1$ when $x = a$, and $0$ otherwise. An MRF prior is placed on ${\bm z}$ to represent the pairwise interactions between groups:
\begin{multline*}
    p({\bf z}_{i,j}|{\bm\lambda})\propto \exp
    \bigg[
    \lambda_1\sum_{s\in\Scal,t\in\Tcal} \mathbbm{1}_{1}\rbr{z_{i,j}^{(s,t)}}
    +
    \lambda_2\sum_{A_s}
    \cbr{
        \mathbbm{1}_{0}\rbr{z_{i,j}^{(s,t)}}
        \mathbbm{1}_{0}\rbr{z_{i,j}^{(s',t')}
    }
    +
        \mathbbm{1}_{1}\rbr{z_{i,j}^{(s,t)}}
        \mathbbm{1}_{1}\rbr{z_{i,j}^{(s',t')}}
    }\\
    +
    \lambda_3\sum_{B_t}
    \cbr{
        \mathbbm{1}_{0}\rbr{z_{i,j}^{(s,t)}}
        \mathbbm{1}_{0}\rbr{z_{i,j}^{(s',t')}
    }
    +
        \mathbbm{1}_{1}\rbr{z_{i,j}^{(s,t)}}
        \mathbbm{1}_{1}\rbr{z_{i,j}^{(s',t')}}
    }
    \bigg].
\end{multline*}
Under this prior distribution, for any tuple $(s,t,s',t')$ in $A_s$ or $B_t$, $z_{i,j}^{(s,t)}$ and $z_{i,j}^{(s',t')}$ have a higher probability to have the same sign. In addition, by varying the values of $\lambda_1, \lambda_2,\lambda_3$ one can weight the importance of individual components, spatially similar components, and temporally similar components,  respectively. \citet{lin2017joint} provided a MATLAB implementation of the joint temporal and spatial estimation. 
}\\\par

{Another example of describing the non-uniform relationships between groups is to build a hierarchical diagram, such as a tree graph. In the Bayesian formulation, we can implement this by constructing a hierarchical prior in the factor form. \citet{oates2014joint} proposed a Structure Learning Trees (SLTs) prior that encodes the hierarchical information among groups. Although the SLTs prior is not originally designed for estimating Gaussian graphical models, the prior can be applied to regularize the structure of the inverse covariance matrices.}
\begin{comment}
\citet{jalali2019bayesian} partitioned precision matrices into the sums of subset-specific matrices and placed specific prior on each subset. It started with decomposing the precision matrix ${\bm\Omega}^{(k)}$ into a collection of power subsets that contains $k$: $${\bm\Omega}^{(k)}=\sum_{r\in v^{(k)}}{\bm \Psi}^{r},\;v^{(k)}=\{r\in\mathcal{P}(K)\mathbin{/}\{0\}:k\in r\},$$
where $\mathcal{P}(K)$ is the power set of K. Then, the subject specific $\mathcal{S}^2$ prior is introduced to the collection $\{{\bm \Psi}^{r}:r\in\mathcal{P}(K)\mathbin{/}\{0\}\}$. The $\mathcal{S}^2$ prior follows a multivariate Gaussian distribution conditioned on a basis of the sparsity patterns. The sparsity pattern follows two binomial distributions with a density threshold to select which distribution is being used; the probability of each edge being selected follows the Bernoulli distribution. At the end, the model parameters are estimated $\{{\bm \Psi}^{r}:r\in\mathcal{P}(K)\mathbin{/}\{0\}\}$ using Gaussian pseudo-likelihood~\citep{khare2015convex}.\\\par
\end{comment}
%The Bayesian approaches do not require specific prior knowledge of the group relations for the data set. Instead, the group relations are estimated with generic priors. This property is particularly desirable because, in most cases, we may not have the structural information for $K$ groups of data.  

\section{Estimating Differential Graphical Models}\label{section:differential_graph}

In contrast to joint estimation, several applications in biomedical research, such as analyzing the gene expression differences in normal cells and cancer cells or differences between the test group and control group, consider the case where $K=2$. Different from the methods in Section~\ref{section:coarse-grained_structure} -- \ref{section:fine-grained_structure}, we will be focusing on finding the ``differences'' rather than finding the ``similarities''. In the high dimensional setting, we assume that the difference of two graphs is sparse. Although the differences between two graphs can be naively estimated by using a joint estimation method first and then finding the difference, procedures that directly estimate the difference are statistically more efficient~\citep{shojaie2021differential}.

In this section, we briefly overview two approaches that estimate graph differences, the direct estimation method and the regularization based approach. For a detailed introduction, see~\citet{shojaie2021differential} for a recent review.

\subsection{Direct Estimation}
The direct approach estimates the difference ${\bm\Delta}={\bm\Omega}^{(1)}-{\bm\Omega}^{(2)}$ without explicitly estimating individual precision matrices ${\bm\Omega}^{(1)}$ and ${\bm\Omega}^{(2)}$. This approach potentially fits a broader class of precision matrices as the individual precision matrices, ${\bm\Omega}^{(1)}$ and ${\bm\Omega}^{(2)}$, need not be sparse, but only the difference ${\bm\Delta}$ is assumed sparse. In addition, jointly estimating ${\bm\Omega}^{(1)}$ and ${\bm\Omega}^{(2)}$ can be challenging when the sparse assumption is violated. \citet{zhao2014direct} directly estimated the difference ${\bm\Delta}$ by solving a constrained minimization problem, noting that, by definition, we have ${\bm\Sigma}^{(1)}{\bm\Delta}{\bm\Sigma}^{(2)}
        -{\bm\Sigma}^{(1)}
        +{\bm\Sigma}^{(2)}={\bf 0}$. 
Consequently, estimating the differential graph ${\bm\Delta}={\bm\Omega}^{(1)}-{\bm\Omega}^{(2)}$ can be achieved by minimizing the following objective:
\begin{align*}
      &\hat{{\bm \Delta}}
      =
      \argmin\|{\bm \Delta}\|_1,
      \quad\text{subject to  }\left|
        \hat{\bm\Sigma}^{(1)}{\bm\Delta}\hat{\bm\Sigma}^{(2)}
        -\hat{\bm\Sigma}^{(1)}
        +\hat{\bm\Sigma}^{(2)}
        \right|_{\infty}\leq\lambda_1,\nonumber
\end{align*}
which is an extension of the CLIME~\citep{cai2011constrained} method. \citet{xu2016semiparametric, yuan2017differential} utilized the symmetry property ${\bm\Sigma}^{(1)}{\bm\Delta}{\bm\Sigma}^{(2)}={\bm\Sigma}^{(2)}{\bm\Delta}{\bm\Sigma}^{(1)}$ and hence $2^{-1}({\bm\Sigma}^{(1)}{\bm\Delta}{\bm\Sigma}^{(2)}+{\bm\Sigma}^{(2)}{\bm\Delta}{\bm\Sigma}^{(1)})-{\bm\Sigma}^{(1)}+{\bm\Sigma}^{(2)}={\bf 0}$.
They defined the objective function as
\begin{align}
&\hat{\bm\Delta}=\argmin \hat\ell({\bm\Delta})+\lambda_1\|{\bm\Delta}\|_1;\label{eq:directdiff}\\
&\hat\ell({\bm\Delta})=\frac{1}{2}\tr({\bm\Delta}\hat{\bm\Sigma}^{(1)}{\bm\Delta}\hat{\bm\Sigma}^{(2)}) - \tr\left\{{\bm\Delta}(\hat{\bm\Sigma}^{(1)}-\hat{\bm\Sigma}^{(2)})\right\},\notag
\end{align}
where the Hessian of the objective with respect to ${\bm\Delta}$ is $(\hat{\bm\Sigma}^{(1)}\otimes \hat{\bm\Sigma}^{(2)}+\hat{\bm\Sigma}^{(2)}\otimes \hat{\bm\Sigma}^{(1)})/2$, which is positive semi-definite. Therefore, $\hat\ell({\bm\Delta})+\lambda_1\|{\bm\Delta}\|_1$ is a convex function with respect to ${\bm\Delta}$, hence a unique minimizer exists.\\\par

{Direct estimation of differential graphs can be extended to other applications as well.  \citet{wang2021direct} proposed a procedure to estimate the differences of two autoregressive models by leveraging the connection between ${\bm\Delta}$ and the difference of a pair transition matrices. \citet{wang2021direct} developed an efficient two-stage estimation procedure by first optimizing~\eqref{eq:directdiff} and then using $\hat{\bm\Delta}$ to solve a regularized least-squared problem in the second stage.} Other recent work extends the direct estimation approach to more expressive structured differential graphs. \citet{na2021estimating} constructed a latent structure estimator where the underlying difference can be formulated as the sum of a low-rank and sparse matrix -- a framework first discussed by~\citet{chandrasekaran2012latent}. \citet{zhao2019direct} extended the direct estimation approach to estimating the differential graph of functional data.

\subsection{Regularization based approach}\label{ssec:diffnetwork_reg}

The node-based learning framework~\citep{mohan2012structured, mohan2014node} assumes that most parts of the graph are shared, and the difference is generated by a node perturbation. When a node is perturbed, the edges connecting this node to others change across $K$ groups. In addition to maximizing the degree of the overlapping structure between groups, the task is to detect perturbed nodes. An intuitive way to look for the perturbed node is to look at the difference of two graphs ${\bm\Omega}^{(1)}-{\bm\Omega}^{(2)}$. When the $j$-th node is being perturbed, the corresponding $j$-th row and $j$-th column of ${\bm\Omega}^{(1)}-{\bm\Omega}^{(2)}$ will have non-zeros, constructing a unique symmetric row-column group. Given that there are several nodes being perturbed, ${\bm\Omega}^{(1)}-{\bm\Omega}^{(2)}$ will be the union of the row-column groups, each stemming from a perturbed node. Using this concept, the Row-Column Overlap Norm (RCON)~\citep{mohan2012structured, mohan2014node} is designed to encourage sparsity in the union of the row-column groups: 
\begin{align*}
    &{\bm P}_{\text{RCON}}({\bm\Omega})=\lambda_1\sum_{k=1}^2\sum_{i,j}|\omega_{i,j}^{(k)}|+\lambda_2 G_q({\bm\Omega}^{(1)}-{\bm\Omega}^{(2)});\\
    &G_q({\bf A})=\min_{{\bf V}:{\bf A}={\bf V}+{\bf V}^\top}f({\bf V}),\quad
    f({\bf V})=\sum_{j=1}^p\|{\bf v}_j\|_q,
\end{align*}
where ${\bf v}_j$ is the $j$-th column of $\bf V$. It is easy to see that when
$q=1$, the RCON penalty is equivalent to the FGL penalty in~\eqref{FGL}. This penalty function simultaneously imposes sparse structure on both the individuals, ${\bm\Omega}^{(1)}$ and ${\bm\Omega}^{(2)}$, and the difference ${\bm\Omega}^{(1)}-{\bm\Omega}^{(2)}$. As mentioned earlier, this method may not work well under the setting that ${\bm\Omega}^{(1)}$ and ${\bm\Omega}^{(2)}$ are not sparse. {Additionally,~\citet{mohan2014node} provided code for estimating
differential graphs.}\\\par

%\subsection{Bayesian Differential Network}\label{ssec:DN_bayesian}
To infer the relative differences between two graphs in a Bayesian formulation, it is intuitive to place a prior on the the differences of two graphs $\theta_{i,j}=E^{(1)}_{i,j}-E^{(1)}_{i,j}$, for every $i<j$. Since the difference $\theta_{i,j}$ is binary, either $0$ (no difference) or $1$ (difference), \citep{mitra2016bayesian} placed a Bernoulli prior distribution ${\rm Ber}(\pi)$ on $\theta_{i,j}$, $i<j$ where $\pi$ follows a Beta distribution, specifying the tendency of being different on two graphs. 

\section{Joint Estimation from Time Series Data}\label{section:time_series_data}

Time-varying graphical models~\citep{Zhou08time, kolar2010estimating,  zhu2018clustered} can be seen as extensions of joint graphical models with groups organized along the time index. The samples are assumed to be generated as ${\bf x}_{i}^{(t)}\sim\mathcal{N}({\bm\mu}^{(t)}, ({\bm\Omega}^{(t)})^{-1})$, $i=1,\hdots,n_t$, where $t=1,\hdots,T$ is the time index. Under such a model, the estimation of time-varying precision matrices and corresponding dynamic networks is challenging as data scarcity is a serious issue: in many problems, we only observe a single sample at  each time point. Therefore, to make the estimation possible, structural assumptions are imposed on how the underlying precision matrices and dynamic networks change over time. Such assumptions control the model complexity and allow for the development of estimation procedures. Examples of structural assumptions on temporal dynamics include piecewise constant and smoothly changing precision matrices, as well as combinations of both. Piece-wise constant structure captures a discrete temporal evolution from one stage to another. For example, the gene regulatory network in a fruit fly can undergo structural changes as the fruit fly develops from an embryo to an adult state. Smooth temporal structure can be used to model the dynamic functional connectivity of brain networks that exhibit smooth temporal evolution from one brain state to another~\citep{shine2016dynamics}. The temporal dynamics of crime rates are often modeled as a combination of smooth dynamics and sudden jumps, where the jumps capture sudden serious crime events. In this section, we will discuss how to apply the FGL penalty and its variants to build a piecewise constant structure. We also introduce a joint estimation framework of multiple autoregressive models to model smooth temporal data.

\subsection{Regularized Estimation}\label{ssec:timeseries_reg}

The FGL penalty has been widely used in time-varying graphical models to model piecewise constant dynamics~\citep{kolar2010estimating, kolar2012estimating, monti2014estimating, hallac2017network}. For instance, Smooth Incremental Graphical Lasso Estimation (SINGLE)~\citep{monti2014estimating} applies the FGL framework to enforce the similarity between consecutive precision matrices:
\begin{align}\label{eq:single}
    {\bf P}_{SINGLE}({\bm \Omega})=\lambda_1\sum_{t=1}^T\sum_{i\neq j}|\omega_{i,j}^{(t)}|+\lambda_2\sum_{t=2}^T\sum_{i\neq j}|{\omega}^{(t)}_{i,j}-{\omega}_{i,j}^{(t-1)}|.
\end{align}
The first term encourages the sparsity of each graph and the second term regularizes the ``jumps'' across time. On the other hand, Group-Fused Graphical Lasso (GFGL)~\citep{gibberd2017regularized} introduces the Frobenius norm as an alternative to encourage neighbouring similarity:
\begin{align*}
    {\bf P}_{GFGL}({\bm \Omega})=\lambda_1\sum_{t=1}^T\sum_{i\neq j}|\omega_{i,j}^{(t)}|+\lambda_2\sum_{t=2}^T\|{\bm\Omega}^{(t)}_{-ii}-{\bf\Omega}_{-ii}^{(t-1)}\|_F,
\end{align*}
where ${\bm\Omega}^{(t)}_{-ii}$ denotes the precision matrix ${\bm\Omega}^{(t)}$ with the diagonal part removed. One may wonder what are the differences in the structure assumptions between the Frobenius norm and the $\ell_1$-norm in~\eqref{eq:single}. The $\ell_1$-norm regularizes individual changes, while the Frobenius norm assumes global changes, implying that several edges within a graph will change simultaneously.\\\par

While the methods introduced in the last paragraph encourage the similarity of two neighboring graphs, the graph that is one-step ahead and that of one-step behind, another idea is to enforce the similarities within multiple steps ahead and behind. This can be done by creating a moving window index set~\citep{yang2020estimating} $\mathcal{N}_w(t)$ for each time point $t=1,\ldots,T$. Consider a window of length $2w$. At every time point $t$, we look at data $w$-steps ahead and $w$-steps behind and hence the index set is $\mathcal{N}_{w}(t)=\{i=1,\ldots,T:|t-i|\leq2w\}$. Note that the index set $\mathcal{N}_{w}(t)$ also includes $t$ itself. Then, we apply the GGL penalty to the components in the index set.

\subsection{Kernel Smoothing Graphical Models}
\label{ssec:ksgm_timeseries}
Another way to construct smoothly varying graphs is by using an autoregressive structure. This model assumes that each data point is a linear combination of previous data points with additional independent noise. Consider the lag-1 case, where ${\bf x}^{(t)}$ is a linear transform  of ${\bf x}^{(t-1)}$ with independent noise ${\bm\varepsilon}^{(t)}\sim\mathcal{N}({\bf 0}, {\bf G}^{(t)})$:
\begin{align*}
&{\bf x}^{(t)} = {\bf A}{\bf x}^{(t-1)} + {\bm\varepsilon}^{(t)},\quad t=1,\ldots, T,
\end{align*}
where ${\bf A}\in\RR^{p\times p}$ is the transition matrix. Consequently, the covariance matrix is smoothly varying along $t$ if ${\bf G}^{(t)}$ is a smooth function of $t$:
\begin{align}\label{autoregressive_covariance}
{\bm\Sigma}^{(t)} = {\bf A}{\bm\Sigma}^{(t-1)}{\bf A}^T + {\bf G}^{(t)},\quad t=1,\ldots, T.
\end{align}
Motivated by this structure, \citet{zhou2010time} proposed a kernel based method to estimate a smooth time-varying covariance structure. First, a weighted sum of the sample covariance matrices $\hat{\bm\Sigma}^{(1)},\ldots, \hat{\bm\Sigma}^{(T)}$ is computed as \[
\hat{\bm S}^{(t)}=\frac{\sum_{s=1}^Tw(s,t)\hat{\bm\Sigma}^{(s)}}{\sum_{s'=1}^Tw(s',t)}
,
\]
where the weights are constructed by a symmetric nonnegative kernel function $K(|s-t|/h)$. This ensures that the estimated covariance is smoothly varying over time. Subsequently, the precision matrix is estimated using the following objective:
\[
    \hat{\bm\Omega}^{(t)}=\argmax\quad n\left[\frac{1}{2}\log\{\det({{\bm\Omega}^{(t)}})\}-\frac{1}{2}\tr(\hat{\bf S}^{(t)}{\bm\Omega}^{(t)})\right]-\lambda_n\sum_{i\neq j}|\omega_{i,j}^{(t)}|,\quad t=1,\ldots,T.
\]

The kernel smoothing method can also be extended to model two-way continuous changes. For instance, the ages of subjects from the fMRI dataset vary across an interval, and one can parametrize the transition matrices as ${\bm A}(u)$ with $u$ taking values in a closed subset of the real line. This model is smooth in two aspects: across the temporal domain and labels (groups).  Hence, we have the following autoregressive model:
\begin{align*}
    &{\bm x}_{i,t}={\bm A}(u_i){\bm x}_{i,t-1}+{\bm\varepsilon}_{i,t},\quad i=1,\ldots,n, \;t=2,\ldots,T.\\
    &{\bm \Sigma}(u)={\bm A}(u){\bm \Sigma}(u){\bm A}(u)^\top+\sigma^2{\bm I}.
\end{align*}
The Kernel-Smoothing Estimator (KSE)~\citep{qiu2016joint} first uses a kernel based estimator for the covariance matrix and then uses the CLIME~\citep{cai2011constrained} method introduced in Section~\ref{ssec:HSN} to recover precision matrices. Consider a set of $n$ data ${\bm Y=\{{\bm y}_1,\hdots,{\bm y}_n\}}$, where ${\bm y}_i=\{{\bm y}_{i,1},\hdots,{\bm y}_{i,T}\}\in\mathbb{R}^{p\times T}$ and with label $u_i\in[0,1]$. The estimated covariance model of the label $u_0\in[0,1]$ is formulated as follows:
\begin{align*}
    &\hat{\bm S}(u_0)=\sum_{i=1}^n w_i(u_0,h)\hat{\bm {\Sigma}}_i;\\
    &w_i(u_0, h):=\frac{c(u_0)}{nh}K\left(\frac{u_i-u_0}{h}\right);\\
    &c(u_0)=\left\{\begin{array}{cc}
         2{\bm I},\;u_0\in\{0, 1\},\\
         1{\bm I},\;u_0\in(0,1),
    \end{array}\right.;\\
    &\hat{\bm\Sigma}_i = \frac{1}{T}\sum_{t=1}^T{\bm y}_{i,t}{\bm y}_{i,t}^\top,
\end{align*}
where $w_i$ is the kernel-based weight with a predefined scale $h$, $K(\cdot)$ is the kernel, $c(u_0)$ determines the boundary value, and $\hat{\bm \Sigma}_i$ is the sampled covariance of the time-series data. After obtaining $\hat{\bm S}(u_0)$, the precision matrix $\hat{\bm\Omega}(u_0)$ is obtained using CLIME in~\eqref{eq:CLIME}:
\begin{align*}
     &\hat{{\bm\Omega}}(u_0)=\argmin \|{\bm\Omega}(u_0)\|_1,\\
    &\text{subject to }|\hat{{\bm S}}(u_0){\bm\Omega}(u_0)-{\bm I}|_\infty\leq\lambda_1.\nonumber 
\end{align*}
Under this framework, the kernel trick is used to capture the assumption that the covariance matrices are smoothly varying across labels. In addition, the Euclidean distance of two labels reflects the similarity of the two groups, capturing the dependence structure. The kernel-based method can be applied to general joint estimation, where the sampled covariance of time-series data is replaced by the sample covariance of data with the same labels.\\\par

\section{Open Problems}\label{section:conclusion}
% 
%As the example we give in Section~\ref{s:intro}, j
%We have presented several approaches for joint estimations of Gaussian graphical models and we will discuss some other open prolems.
Existing and emerging biological data and applications will require novel approaches to joint graphical models. We discuss some of these emerging applications briefly. Joint estimation of functional connectivity networks across multiple subjects allows scaling of the effective sample size and computation of estimates that are more robust to outliers. The joint estimators of brain connectivity networks could be applied to task-based fMRI scans to study group dynamic functional connectivity patterns~\citep{andersen2018bayesian, calhoun2014chronnectome, gonzalez2018task}. While this manuscript is focused on joint estimation with the same set of nodes, one potential direction is to extend it to multiple sources, i.e.,  multimodal data. Recent technologies~\citep{huster2012methods, abreu2018eeg} have demonstrated the availability of conducting concurrent measurements of EEG and fMRI signals, allowing the estimation of multiple sources possible in the future. While EEG has a higher temporal resolution and fMRI features a higher spatial resolution, we believe that joint estimation with multiple sources could compensate for the limits of the measurement techniques and provide better estimation results. Some recent work~\citep{lock2013joint,li2021integrative} has developed methodologies to integrate data from different modalities, however, joint estimation of graphical models from multimodal data is still an open problem.
\\\par

Approaches for the estimation of the joint graphical models presented in this survey largely rely on penalized estimation, where the penalty biases the estimates towards the assumed structure. Quantifying statistical uncertainty about the model parameters, that is, performing hypothesis tests  and constructing confidence intervals, is challenging when penalized estimators are used due to the induced bias and model selection  that is implicitly performed. There has been recent work on statistical inference for low-dimensional parameters in graphical models \citep{Ren2013Asymptotic, Jankova2014Confidence, Jankova2017Honest,Barber2015ROCKET,Wang2016Inference, Yu2016Statistical, Yu2019Simultaneous} based on the $\ell_1$-penalized estimator in the first stage. However, these approaches were developed only in the setting where  parameters of one graph are being inferred. In contrast, work on statistical inference for joint graphical models is much more sparse. \citet{Xia2015Testing, Belilovsky2016Testing2, Liu2017Structural, kim2021two} developed techniques for statistical inference in differential graphical models, while \citet{Wang2014Inference, Lu2015Posta, Wang2020Statistical} focused on graphical models for time series data. {
\citet{wang2021joint} developed a hierarchical testing procedure for joint inferences of multiple graphs on Hawkes processes, albeit in non-Gaussian settings.} Developing the corresponding inferential techniques for estimators obtained using coarse-grained and fine-grained penalties is an interesting area open for future research.

\section{Conclusion}

This manuscript has introduced joint Gaussian graphical model estimation methods for joint data with shared structure across multiple groups. In particular, we have considered several examples of extending classical statistical inference methods to joint estimation settings, including the MLE based estimator, neighborhood regression, and the CLIME estimator. We have discussed several methods that exploit coarse-grained structures using a global regularization method that encourages a shared coarse-grained structure across all groups. In contrast, the fine-grained structural regularization methods further partition the groups into subgroups per node, encouraging local shared regularity. With two groups, differential graphs are often a highly effective approach. We have also discussed the applications of joint estimation techniques to the estimation of graphical models from time-series data.
%The presented simulation experiments demonstrate that joint estimation 
%can often recover the graphs more effectively than separate estimation for
%high dimensional data. 
%Furthermore, we verify the characteristics of fused graphical lasso and group graphical regularizers for two data generation processes. When the data 
%generation process has a shared inverse covariance structure with different values,
%group graphical lasso methods may be more effective. On the other hand, when 
%the data generation process features a shared inverse covariance structure with the same %values,
%the fused graphical lasso penalty function may perform better.

\section*{Funding Information}
O.Koyejo acknowledges partial funding from a C3.ai Digital Transformation Institute Award, a Jump Arches Award, and an Strategic Research Initiatives award from the University of Illinois at Urbana-Champaign. K.Tsai acknowledges funding from National Science Foundation Graduate Research Fellowships Program. Other authors have no relevant financial or nonfinancial interests to disclose. This work was also funded in part by the following grants: NSF III 2046795 and IIS 1909577, along with computational resources donated by Microsoft Azure.
%\section*{Research Resources}
%\section*{Acknowledgments}

\section*{Further Reading}
Recent developments of joint statistical inference are primarily focused on Gaussian graphical models. Other types of graphical models, including discrete graphical models~\citep{drton2008binary, drton2009discrete}, semiparametric/nonparametric graphical models~\citep{liu2012high,sun2015learning}, and latent graphical models~\citep{chandrasekaran2012latent}, have been well studied for single graph estimation. While such models have broad applications, joint estimation in these models
is less studied. 
\newpage
\medskip
\bibliographystyle{apalike}
%\printbibliography
%\bibliographystyle{apacite}
\bibliography{tmp,ref}

\end{document}